\documentclass[12pt,onecolumn]{IEEEtran}

\usepackage{cite}
\usepackage{graphicx}
\usepackage{amsfonts,amsmath,amssymb}
\usepackage{url}
\usepackage{bm}
\usepackage{bbm}
\usepackage{subfigure}
\usepackage{stfloats}

\usepackage{algorithmic}
\usepackage{algorithm}

\usepackage{graphicx,booktabs,color}

\newtheorem{theorem}{\textbf{Theorem}}
\newtheorem{proposition}{\textbf{Proposition}}
\newtheorem{lemma}{\textbf{Lemma}}

\begin{document}

\title{Beamforming Design and Power Allocation for Secure Transmission with NOMA}

\author{Youhong Feng, \IEEEmembership{Student Member, IEEE,} Shihao Yan, \IEEEmembership{Member, IEEE,}  Zhen Yang,\\ Nan Yang, \IEEEmembership{Member, IEEE,} and Jinhong Yuan, \IEEEmembership{Fellow, IEEE}

\thanks{This work was partially supported by the National Natural Science Foundation of China (No.61772287, 61671252).}
\thanks{Y. Feng and Z. Yang are with the Key Laboratory of Ministry of Education in Broadband Wireless Communication and Sensor Network Technology, Nanjing University of Posts and Telecommunications, Nanjing 210003, China (e-mails: \{2013010213, yangz\}@njupt.edu.cn). Y. Feng is also with the College of Physics and Electronic information Engineering, Anhui Normal
University, Wuhu 241000, China,  and the Research School of Engineering, Australian National University, Canberra, ACT, Australia. S. Yan is with the School of Engineering, Macquarie University, NSW 2109, Australia (e-mail:shihao.yan@mq.edu.au ). N. Yang is with the Research School of Engineering, Australian National University, Canberra, ACT 2600, Australia (email: nan.yang@anu.edu.au). J. Yuan is with the School of Electrical Engineering and Telecommunications, The University of New South Wales, Sydney, NSW 2052, Australia (e-mail: j.yuan@unsw.edu.au).}}


\markboth{Submitted to IEEE TRANSACTIONS ON Wireless Communications}{Feng \MakeLowercase{\textit{et al.}}: Beamforming Design and Power Allocation for Secure Transmission with NOMA}

\maketitle

\vspace{-2.5cm}

\begin{abstract}
In this work, we propose a novel beamforming design to enhance physical layer security of a non-orthogonal multiple access (NOMA) system with the aid of artificial noise (AN). The proposed design uses two scalars to balance the useful signal strength and interference at the strong and weak users, which is a generalized version of the existing beamforming designs in the context of physical layer security for NOMA. We determine the optimal power allocation among useful signals and AN together with the two optimal beamforming scalars in order to maximize the secrecy sum rate (SSR). Our asymptotic analysis in the high signal-to-noise ratio regime provides an efficient and near-optimal solution to optimizing the beamforming scalars and power allocation coefficients. Our analysis indicates that it is not optimal to form a beam towards either the strong user or the weak user in NOMA systems for security enhancement. In addition, the asymptotically optimal power allocation informs that, as the transmit power increases, more power should be allocated to the weak user or AN signals, while the power allocated to the strong user keeps constant. Our examination shows that the proposed novel beamforming design can significantly outperform two benchmark schemes.
\end{abstract}

\vspace{-0.5cm}

\IEEEpeerreviewmaketitle

\section{Introduction}
Non-orthogonal  multiple access (NOMA), as a potentially promising technique to significantly boost the system spectral efficiency in the fifth-generation (5G) and beyond wireless networks, has attracted an increasing amount of research effort~\cite{Y.Saito2013,Z.Ding2017JSAC,L.Dai2015,Y. Liu2017-1}. Different from the conventional orthogonal multiple access (OMA) techniques, such as frequency division multiple access, time division orthogonal multiple access, and code division multiple access, NOMA  can exploit the power domain to serve multiple users simultaneously in the same resource block (i.e., time/frequency/code), in which successive interference cancellation (SIC) is widely applied. Motivated by the improved spectral efficiency provided by NOMA, different issues in NOMA systems have been addressed in the literature (e.g., \cite{A.Benjebbovu2013,J. Choi2016,Y. Sun2017,X.Sun201802,X.Sun201801}).
For example, \cite{A.Benjebbovu2013} focused on a downlink NOMA system, where the authors considered user
pairing and transmit power allocation to enhance the performance of NOMA. In \cite{J. Choi2016}, an optimal transmit power allocation scheme  was proposed in multiple-input multiple-output (MIMO) NOMA systems in order to maximize the sum rate of two paired users subject to some specific constraints. In addition, the authors of \cite{Y. Sun2017} proposed a joint subcarrier and power allocation scheme to maximize the weighted sum rate of a NOMA system. Considering delay constraint, the authors of \cite{X.Sun201802} tackled the maximization of the effective throughput in the context of NOMA for short-packet communications, which shows that NOMA can aid to achieve low-latency communications.

Wireless communication security is another critical issue of growing importance in 5G and beyond wireless networks, since there is an increasing amount of confidential information (e.g., credit card information) that is transferred over the air. Physical layer security, as a complementary and alternative cryptographic method to defend against eavesdroppers, exploits the
inherent properties (e.g., randomness) of the wireless medium to achieve the ever-lasting and information-theoretic secrecy (e.g., \cite{yang2015safeguarding,zou2016,YouhongFeng2017,Y.Hong2016,Y. Feng201702,T.-X.Zheng2014,S.Yan2015,N.Yang2016}). In this context, MIMO architectures (e.g., \cite{Y.Hong2016,T.-X.Zheng2014}) and artificial-noise (AN)-aided secure transmissions (e.g., \cite{YouhongFeng2017,Y. Feng201702,yang2015safeguarding,Y.Hong2016,N.Yang2016}) have been widely adopted to enhance the secrecy performance of wireless communications. Against this background, physical layer security in NOMA systems has been partially addressed \cite{Z.Ding2017TC,Z.Qin2016,Y.Zhang2016,M.Tian2017sp,M. Jiang2017sp,Y.Zhang2017,B.He2017,Y.Liu2017,F. Zhou2018,L.Lv2018}. For example, in \cite{Y.Zhang2016} the authors considered physical layer security in a single-input single-output (SISO) NOMA system and proposed an optimal power allocation policy for maximizing the secrecy sum rate (SSR) of all users subject to their predefined quality of service (QoS) requirements. In \cite{F. Zhou2018}, the authors focused on the transmission power minimization in a multiple-input single-output (MISO) NOMA cognitive radio network in the presence of multiple single-antenna eavesdroppers. Considering a cell-edge user (i.e., the weak user) as a potential eavesdropper to an entrusted central user (i.e., the strong user), the maximization of the secrecy rate of the central user subject to a transmit power constraint and a transmission rate requirement at the cell-edge user is tackled in \cite{M. Jiang2017sp}. In \cite{M.Tian2017sp}, the authors focused on the SSR optimization problem for a downlink MIMO NOMA network subject to successful SIC  and transmit power constraints, in which the nonconvex maximization of the SSR was transfer to a biconvex problem that was solved by alternating optimization method.

In the considered NOMA systems of \cite{M.Tian2017sp,M. Jiang2017sp,F. Zhou2018}, either the perfect knowledge on the eavesdropper's instantaneous channel state information (CSI) or a bounded error model on the the eavesdropper's instantaneous CSI was considered. Such CSI information may not be achievable in some specific application scenarios of NOMA, in which the eavesdropper is not an internal user or an active receiver. As such, the assumption that only the statistical information on the eavesdropper's CSI (e.g., a passive eavesdropping scenario) was widely used in the context of physical layer security for NOMA (e.g., \cite{Y.Liu2017,B.He2017,L.Lv2018}).
Specifically,  \cite{B.He2017} proposed a NOMA scheme that maximizes the minimum confidential information rate under the secrecy outage probability (SOP) and transmit power constraints. Inspired by the enhanced secrecy performance achieved by AN-aided transmission strategies, \cite{Y.Liu2017} and \cite{L.Lv2018} considered AN-aided secure beamforming (SBF) strategies to protect the confidential information of legitimate users for MISO NOMA systems. More specifically, the authors of \cite{Y.Liu2017} considered large-scale networks with randomly deployed legitimate users and eavesdroppers, where the exact and asymptotic expressions for the SOP were derived. The imperfect SIC was considered in \cite{L.Lv2018}, where the SOPs of the legitimate users were obtained in closed-form expressions.

With regard to the SBF design in NOMA systems, in \cite{Y.Liu2017} a maximum ratio transmission (MRT) strategy was adopted, i.e., $\mathbf{v}_1 = \mathbf{h}_1/|\mathbf{h}_1|$ and $\mathbf{v}_2 = \mathbf{h}_2/|\mathbf{h}_2|$, where $\mathbf{v}_1$ and $\mathbf{v}_2$ are the beamforming vectors used to transmit useful signals to the weak user (User 1) and the strong user (User 2), respectively, while $\mathbf{h}_1$ and $\mathbf{h}_2$ are the channel vectors from the transmitter to User 1 and User 2, respectively. In this MRT strategy, the signal strengths of $s_1$ and $s_2$ are maximized at User 1 and User 2, respectively.
As clarified in \cite{L.Lv2018}, this MRT strategy may not guarantee perfect SIC at User 2 (i.e., the strong user), since the interference caused by $s_2$ is also maximized when User 2 decodes $s_1$ in order to conduct SIC. Thus, in \cite{L.Lv2018} the SBF is designed such that $\mathbf{v}_1 = \mathbf{v}_2 = \mathbf{h}_2/|\mathbf{h}_2|$, i.e., the signal strengths of both $s_1$ and $s_2$ are maximized at User 2. In the SBF designs proposed by \cite{Y.Liu2017} and \cite{L.Lv2018}, we observe that the freedom of balancing the useful signal strength and interference is lost, i.e., the useful signal strengths and interference are either minimized or maximized. We note that this freedom can potentially enhance the achieved physical layer security in NOMA systems, since the useful signal $s_1$ should be decoded at both User 1 and User 2, while the useful signal $s_2$ causes interference at both User 1 and User 2 (but $s_2$ is only decoded at User 2). To design a SBF, for which the freedom to balance useful signal strengths and interference can be achieved in order to improve the secrecy performance of NOMA systems, motivates this work and our main contributions are summarized as below.
\begin{itemize}
\item We propose a novel hybrid SBF scheme in a NOMA system, which can balance the useful signal strengths and interference at both User 1 and User 2 in order to enhance physical layer security. Specifically, in our proposed scheme $\mathbf{v}_1$ is a linear function of $\mathbf{h}_1$ and $\mathbf{h}_2$, which is determined by a parameter $\beta_1$, and $\mathbf{v}_2$ is a linear function of $\mathbf{h}_1$ and $\mathbf{e}$, which is determined by another parameter $\beta_2$, where $\mathbf{e}$ is random vector that does not align with $\mathbf{h}_1$ or $\mathbf{h}_2$. In this scheme, AN is also used to further enhance the secrecy performance of NOMA systems and thus we refer to this scheme as the NOMA-HB-AN scheme, in which the beams used to transmit $s_1$ and $s_2$ can be flexibly controlled to be in any direction (e.g., maybe towards neither User 1 nor User 2). We note that the proposed NOMA-HB-AN scheme is a generalized version of the SBF designs proposed in \cite{Y.Liu2017} and \cite{L.Lv2018}.

\item In order to maximize the benefits of the proposed NOMA-HB-AN scheme, we tackle the optimization of the two governing parameters $\beta_1$ and $\beta_2$ together with the optimal power allocation among $s_1$, $s_2$, and the AN signals, aiming to maximize the SSR subject to specific QoS constraints at the two legitimate users. Considering a larger number of transmit antennas, we first determine the optimal power allocation for given $\beta_1$ and $\beta_2$, in which the power allocation coefficients for $s_1$ and $s_2$ are analytically derived as functions of the power allocation coefficient for AN signals. This leads to that the optimal power allocation can be achieved with the aid of a one-dimensional  numerical search. Our examination shows that the proposed NOMA-HB-AN scheme can significantly outperform the SBF design with $\mathbf{v}_1 = \mathbf{v}_2 = \mathbf{h}_2/|\mathbf{h}_2|$ proposed in \cite{L.Lv2018}.

\item To gain further insights on the proposed scheme, we consider the joint optimization of $\beta_1$ and $\beta_2$ together with power allocation in the regime of high signal-to-noise (SNR) ratio. Particularly, we derive the power allocation coefficients for $s_1$, $s_2$, and the AN signals in closed-form expressions in the high-SNR regime, based on which the optimization of $\beta_1$ and $\beta_2$ can be efficiently achieved by another one-dimensional numerical search. Our numerical examination shows that the achieved optimal $\beta_1$, $\beta_2$, and power allocation in the high-SNR regime can precisely approximate those achieved for arbitrary SNRs, in terms of achieving similar maximum SSRs. This indicates that our proposed SBF design can be efficiently optimized and the associated complexity increase is negligible. Our results also show that, when the eavesdropper's channel quality is not that high, the design of SBF is more important than whether to use AN, which is confirmed by our observation that the proposed SBF design without AN can even outperform the design with $\mathbf{v}_1 = \mathbf{v}_2 = \mathbf{h}_2/|\mathbf{h}_2|$ and AN.
\end{itemize}

The remainder of this paper is organized as follows. In Section II, the system model and the hybrid SBF are presented. The maximization of the SSR under the QoS constraints at the two legitimate users is formulated in Section III. The solution to the SSR maximization problem are provided in Section IV, where the scenarios with arbitrary and high SNRs are considered.
Numerical results are provided in Section V to offer valuable insights on the secrecy performance of the proposed scheme compared with two benchmark schemes. Conclusions are drawn in Section VI.

\textit{Notation:} Scalar variables are denoted by italic symbols; Vectors and matrices are denoted by lower-case and upper-case boldface symbols, respectively;  $\mathbf{A}^{{H}}$ denotes the Hermitian
(conjugate) transpose of a matrix $\mathbf{A}$; $\mathbf{I}_{{K}}$ represents the $K\times K$ identity matrix; $E[x]$
denote the mean  of the random
variable $x$; $x\sim \textit{CN}(\mu, \sigma^{2})$  denotes a circularly  symmetric complex Gaussian random variable $x$ with mean $\mu$ and covariance $\sigma^{2}$.

\section{System Model}

\begin{figure}[!t]
\centering
 \includegraphics[width=3.4in, height=2.7in]{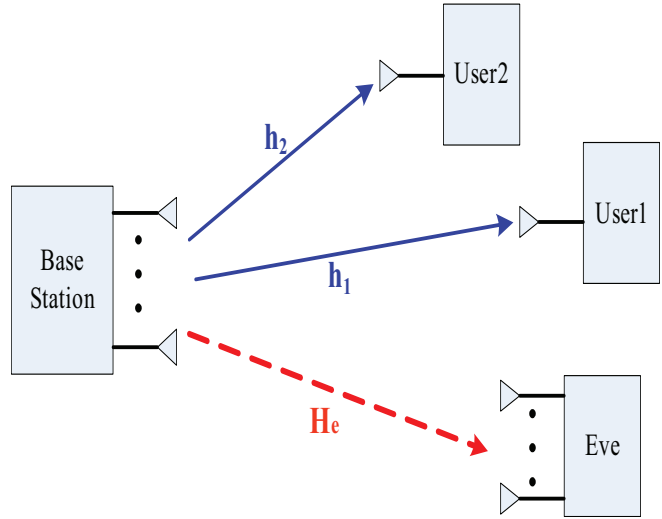}
\caption{Illustration of a downlink MISOME NOMA system, where the BS is equipped with $N$ antennas,  each of User 1 and User 2 is equipped with a single antenna, and the eavesdropper is equipped with $K$ antenna.}
\label{fig_sim2}
\end{figure}


We consider the secure transmission using NOMA from a base station (BS) to two legitimate users in the presence of a multi-antenna eavesdropper (Eve). The BS is equipped with $N$ antennas, each of the legitimate users (i.e., User 1 and User 2) is equipped with a single antenna, and Eve is equipped with $K$  antennas. As such, we refer to the considered system as a MISOME NOMA system.
We assume that $N$ is large, and in particular is much larger than $K$. The channel vector from the BS to the legitimate user $m\in\{1,2\}$ is denoted by $\mathbf{h}_{m}\in \mathbb{C}^{1 \times N }$, of which the entries are independent and identically distributed (i.i.d.) circularly-symmetric complex Gaussian random variables with zero-mean and variance $\delta_{m}^{2}$. The channel matrix from the BS to Eve is denoted by $\mathbf{H}_{e}\in\mathbb{C}^{K \times N}$, where $\mathbf{h}_{e,k}\triangleq\mathbf{H}_{e}(k,:)$ and $\mathbf{h}_{e,k}  \in \mathbb{C}^{1 \times N}$ is an $1\times N$ channel vector from the BS to the $k$-th receive antenna at Eve and $\mathbf{h}_{e,k}$ is i.i.d. circularly-symmetric complex Gaussian entries with zero-mean and variance $\delta_{e}^{2}$.

In this work,  we assume that the CSI of all the legitimate channels (i.e., $\mathbf{h}_{m}$) is known at the BS, while only the statistical CSI of the Eve's channel (i.e., the statistical information on $\mathbf{H}_{e}$) is available. It is a very generic assumption that the statistical CSI of the Eve's channel is known, which has been widely adopted in the literature of physical layer security~\cite{L.Lv2018,L.Fan2016,S.Yan2016}. Without loss
of generality, we assume that the legitimate channel gains
are sorted in ascending order\cite{Y.Zhang2017,B.He2017}, i.e., $0<\|\mathbf{h}_{1}\|^{2}\leq \|\mathbf{h}_{2}\|^{2}$.

\subsection{Secure Transmission with NOMA and Artificial Noise}

We next detail the secure transmission using NOMA with AN in our considered system model. Specifically, the BS transmits two information signals,  $s_{1}$ and $s_{2}$, in conjunction with an $(N-2)\times 1$ AN vector $\mathbf{s}_{N}$ to its corresponding receivers, where $s_{m}$ is the information signal dedicated for the $m$-th user. The variance of $s_{m}$ is denoted by $\chi_{m}$ and the total transmit power is denoted by $P$. We denote $\phi_{m}$ as the power allocation coefficient to $s_{m}$, where $0<\phi_{m}\leq 1$, which determines the fraction of the total transmit power allocated to $s_{m}$ such that $\chi_{m}=\phi_{m}P$. Since the BS does not know $\mathbf{H}_{e}$, it equally distributes the AN transmit power to each entry of $\mathbf{s}_{N}$ and thus the variance of each entry of $\mathbf{s}_{N}$ is the same, which is denoted by $\chi_{N}$. Then the BS transmits $\mathbf{s}_{N}$ in the null space of the channel from the BS to the two users, $\mathbf{H}\triangleq[\mathbf{h}_{1}^{H},\mathbf{h}_{2}^{H}]$, such that $\mathbf{s}_{N}$ leads to interference at Eve but not at the two legitimate users. As such, we know that all the remaining transmit power (excluding the power allocated to $s_{1}$ and $s_{2}$) should be used
to transmit $\mathbf{s}_{N}$, such that we have $\chi_{N}=\phi_{e}P/(N-2)$ with $\phi_{e}=1-\phi_{1} - \phi_2$. To transmit $s_{m}$ and $\mathbf{s}_{N}$, the BS has to design an $N\times N$ beamforming matrix $\mathbf{V}$ given by
\begin{align}\label{BF-V}
      \mathbf{V}=\left[\mathbf{v}_{1},\mathbf{v}_{2},\mathbf{V}_{N}\right],
\end{align}
where we recall that $\mathbf{v}_{1}$ and $\mathbf{v}_{2}$ are the beamforming vectors used to transmit $s_{1}$ and $s_{2}$, respectively, and $\mathbf{V}_{N}$ is the unitary beamforming matrix used to transmit $\mathbf{s}_{N}$. In this work, we adopt specific structures for $\mathbf{v}_{1}$ and $\mathbf{v}_{2}$ given as
\begin{align}
\mathbf{v}_{1}&=\frac{\sqrt{\beta_{1}}\mathbf{\hat{h}}_{1} + \sqrt{(1-\beta_{1})}\mathbf{\hat{h}}_{2}}{\|\sqrt{\beta_{1}}\mathbf{\hat{h}}_{1} +\sqrt{ (1-\beta_{1})}\mathbf{\hat{h}}_{2}\|},\label{beamforming-1}\\
\mathbf{v}_{2}&=\frac{\sqrt{\beta_{2}}\mathbf{\hat{h}}_{2} + \sqrt{(1-\beta_{2})}\mathbf{\hat{e}}}{\|\sqrt{\beta_{2}}\mathbf{\hat{h}}_{2} + \sqrt{(1-\beta_{2})}\mathbf{\hat{e}}\|},\label{beamforming-2}
\end{align}
where $\beta_1$ and $\beta_2$ are design parameters to be determined later, $\mathbf{\hat{h}}_{i}=\frac{\mathbf{h}_{i}}{\|\mathbf{{h}}_{i}\|}$, $\mathbf{\hat{e}}=\frac{\mathbf{{e}}}{\|\mathbf{{e}}\|}$, and $\mathbf{e} \in \mathbb{C}^{1 \times N}$ is a random vector that does not align with $\mathbf{h}_1$ or $\mathbf{h}_2$. The design of $\mathbf{v}_{1}$ originates from the fact that the information signal $s_{1}$ need to be decoded by both User 1 and User 2 (User 2 decodes $s_{1}$ by performing SIC) and the design of $\mathbf{v}_{2}$ originates from that only User 2 decodes $s_{2}$ while $s_{2}$ causes interference at User 1 for decoding $s_{1}$. We note that the proposed $\mathbf{v}_{1}$ and $\mathbf{v}_{2}$ are generalizations of the beamforming vectors adopted in existing works (e.g., \cite{L.Lv2018,Y.Liu2017}) and thus they are expected to achieve better system performance with optimized $\beta_1$ and $\beta_2$, which will be confirmed by our examination in this work.
Using $\mathbf{\textbf{V}}$, the transmitted signal vector at the BS is given by
\begin{align}\label{Sig-S}
  \mathbf{s}&=\mathbf{V}\left[
 \begin{array}{ccc}
 s_{1} \\
      s_{2} \\
 \mathbf{s}_{N}\\
 \end{array}
\right]
=\mathbf{v}_{1}s_{1}+\mathbf{v}_{2}s_{2}+\mathbf{V}_{N}\mathbf{s}_{N}.
\end{align}
Therefore, the received signal at the $m$-th user is given by
\begin{align}\label{Rec-user-k}
 y_{m}&=\mathbf{h}_{m}\mathbf{s}+n_{m}\notag\\
&=\mathbf{h}_{m}\sum_{i=1}^{2}\mathbf{v}_{i}s_{i}+\mathbf{h}_{m}\mathbf{V}_{N}\mathbf{s}_{N}+n_{m}\notag\\
&=\mathbf{h}_{m}\sum_{i=1}^{2}\mathbf{v}_{i}s_{i}+n_{m},
\end{align}
where $n_{m}$ satisfying $E[n_{m}n_{m}^{H}]=\sigma_{m}^{2}$ is the additive white Gaussian noise (AWGN) at the $m$-th user and $\mathbf{h}_{m}\mathbf{V}_{N} = \mathbf{0}$ is applied, since $\mathbf{s}_{N}$ is transmitted in the null space of $\mathbf{H}$.
Likewise, the received signal vector at Eve is given by
\begin{align}\label{Rec-Eave-E}
\mathbf{y}_{e}&=\mathbf{H}_{e}\mathbf{s}+\mathbf{n}_{e}=\mathbf{H}_{e}\sum_{i=1}^{2}\mathbf{v}_{i}s_{i}+\mathbf{H}_{e}\mathbf{V}_{N}\mathbf{s}_{N}+\mathbf{n}_{e},
\end{align}
where $\mathbf{n}_{e}$ satisfying $E[\mathbf{n}_{e}\mathbf{n}_{e}^{H}]=\mathbf{\sigma}_{e}^{2}\mathbf{I}_{K}$ is the AWGN vector at Eve.

\subsection{Performance Metric for NOMA with AN}

According to the principle of NOMA, the user with the better channel condition (i.e., User 2) firstly decodes the signal of the other user (i.e., User 1) and then successively subtracts the interference caused by this signal from its received signal before decoding its own information. The weak user, User 1 directly decodes its own information by treating User 2's signal as interference~\cite{Z.Ding2014}. As such, the maximum achievable rate of $s_1$ is given by \cite{L.Lv2018,Z.Chen2016}
 \begin{align}\label{Rate-user-1}
R_{u1}=\log_{2}\left(1+\gamma_{u1}\right),
\end{align}
where
\begin{align}\label{Rate-user-2-xishu}
\gamma_{u1}=\min\left(\frac{\phi_{1}P|\mathbf{h}_{1}\textbf{v}_{1}|^{2}}{\phi_{2}P|\mathbf{h}_{1}\mathbf{v}_{2}|^{2}+\sigma_{1}^{2}},
\frac{\phi_{1}P|\mathbf{h}_{2}\textbf{v}_{1}|^{2}}{\phi_{2}P|\mathbf{h}_{2}\mathbf{v}_{2}|^{2}+\sigma_{2}^{2}}\right),
\end{align}
while the first and second terms on the right-hand side of \eqref{Rate-user-2-xishu} denote the received  SINR for decoding User 1's signal $s_{1}$ at User 1 and User 2, respectively. We note that the ``$\min$'' function used in \eqref{Rate-user-2-xishu} comes from the assumption that perfect SIC is guaranteed at User 2, which is assumed in this work. With SIC, User 2 decodes its information signal without interference and thus, the maximum achievable rate of $s_2$ is given by
\begin{align}\label{Rate-user-2}
R_{u2}=\log_{2}\left(1+\gamma_{u2}\right),
\end{align}
where $\gamma_{u2} = {\phi_{2}P|\mathbf{h}_{2}\textbf{v}_{2}|^{2}}/{\sigma_{2}^{2}}$ denotes the SNR for decoding User 2's signal at User 2.

In this work, we consider a worst-case scenario, where Eve has already decoded the information signal for User 1 in order to conduct SIC before it attempts to decode the information for User 2, which is exactly the same as the decoding procedure at User 2 (i.e., the strong user). The worst-case assumption has been widely adopted in designing and analyzing the NOMA transmission schemes with physical layer security (e.g., \cite{Y.Zhang2016,B.He2017}). As such, the maximum achievable rate of $s_1$ at Eve is given by \cite{S. Goel2008,S-HTsai2014}
\begin{align}\label{Rate-Eva-1}
&R_{e1}=\log_2\det\left(\sigma_{e}^{2}\mathbf{I}_{K}+\frac{\phi_{1}P\mathbf{H}_{e}\mathbf{v}_{1}(\mathbf{H}_{e}\mathbf{v}_{1})^{H}}{\phi_{2}P\mathbf{H}_{e}\mathbf{v}_{2}(\mathbf{H}_{e}\mathbf{v}_{2})^{H}\!+\!\frac{\phi_eP}{N\!-\!2}\mathbf{H}_{e}\mathbf{V}_{N}(\mathbf{H}_{e}\mathbf{V}_{N})^{H}\!+\!\sigma_{e}^{2}\mathbf{I}_{K}}\right).
\end{align}
Likewise, the maximum achievable rate of $s_2$ at Eve is given by \cite{S. Goel2008,S-HTsai2014}
\begin{align}\label{Rate-Eva-2}
&R_{e2}=\log_2\det\left(\sigma_{e}^{2}\mathbf{I}_{K}+\frac{\phi_{2}P\mathbf{H}_{e}\mathbf{v}_{2}(\mathbf{H}_{e}\mathbf{v}_{2})^{H}}{\frac{\phi_eP}{N\!-\!2}\mathbf{H}_{e}\mathbf{V}_{N}(\mathbf{H}_{e}\mathbf{V}_{N})^{H}+\sigma_{e}^{2}\mathbf{I}_{K}}\right).
 \end{align}

We denote the achievable secrecy rate of $s_m$ and the SSR as $R_{sm}$ and $R_s$, respectively. Therefore, we have
\begin{align}\label{SSR-1}
R_{s}&=\sum_{m=1}^{2}R_{sm}=\sum_{m=1}^{2}\left[R_{um}-R_{em}\right]^{+},
   \end{align}
where $[x]^{+}\triangleq\max(0,x)$. We note that the SSR normally cannot be adopted as the secrecy performance metric in the passive eavesdropping scenario, where the instantaneous CSI of the Eve's channel is unknown. However, as the number of transmit antennas (i.e., $N$) approaches infinity (which is the focus of this work), the CSI of the Eve's channel is asymptotically achievable. As such, following \cite{N.Li2015,N.Li2016} in this work we adopt the SSR as our performance metric under some specific constraints.

\section{Optimization Framework with A Sufficient Large Number of Transmit Antennas}

In this section, we first present the adopted optimization framework. By considering a sufficiently large number of transmit antennas, we conduct new analysis to simplify the objective function and the corresponding constraints.

\subsection{Optimization Framework}

In this work, following \cite{Y.Zhang2016,Y.Zhang2017} we aim to maximize the SSR (i.e., $R_s$) subject to some constraints on the maximum achievable rates of $s_1$ and $s_2$ (i.e., $R_{u1}$ and $R_{u2}$). Specifically, the focused optimization problem can be written as
\begin{align}\label{YOHU-1}
&\textbf{P1}:\operatorname*{max}\limits_{\phi_{1},\phi_{2}, \phi_e, \beta_1, \beta_2} R_{s}\,\,\\
&{\rm  s.t.}\hspace{1mm} R_{um} \geq Q_m, ~m \in \{1, 2\},\label{YOHU-1a} \\
&\hspace{6mm} \phi_1 + \phi_2 + \phi_e = 1, \label{YOHU-1b} \\
&\hspace{6mm} 0 \leq \beta_m \leq 1, ~m \in \{1, 2\}, \label{YOHU-1c}
\end{align}
where $Q_{m}$ denotes the minimum codeword rate required by the $m$-th legitimate user. We note that the constraint given in \eqref{YOHU-1a} can be justified by the fact that secure transmission is only considered when the QoS without security at the $m$-th user is above a specific threshold \cite{Y.Zhang2016,Y.Zhang2017,Y. Sun2018}. We also note that in the optimization problem given in \eqref{YOHU-1} we have $0 < \phi_m < 1$ and $0 \leq \phi_e < 1$.

Due to the constraint given in \eqref{YOHU-1a}, there exists a minimum transmit power, denoted by $P_{\min}$, that guarantees the feasibility of the optimization problem $\textbf{P1}$. In other words, the optimization problem $\textbf{P1}$ is feasible only when $P\geq P_{\min}$. The value of $P_{\min}$ can be determined following the method given in \cite{Y.Zhang2016} and thus in this work we assume that this feasible condition is always guaranteed.

\subsection{SINR of $s_1$ and  Constraint $R_{um} \geq Q_m$}

In this subsection, we present the determined expression (without ``min'') for the SINR of $s_1$ in the following lemma  to facilitate solving the optimization problem $\textbf{P1}$, based on which we also transfer the constraint $R_{um} \geq Q_m$ into a specific constraint on $\phi_m$.

 \begin{lemma}\label{Lemma1}
In the solution to the optimization problem $\textbf{P1}$, the achieved SINR for $s_{1}$ is given by
\begin{align}\label{C1-capacity}
\gamma_{u1}=\frac{\phi_{1}P|\mathbf{h}_{1}\textbf{v}_{1}|^{2}}{\phi_{2}P|\mathbf{h}_{1}\mathbf{v}_{2}|^{2}+\sigma_{1}^{2}}
= \frac{\phi_{1}P|\mathbf{h}_{2}\textbf{v}_{1}|^{2}}{\phi_{2}P|\mathbf{h}_{2}\mathbf{v}_{2}|^{2}+\sigma_{2}^{2}}.
\end{align}
\end{lemma}
\begin{IEEEproof}
Following \eqref{Rate-user-2-xishu}, in order to prove this lemma we only have to prove that
\begin{align}\label{Rate-user-2-xishu-2}
\frac{\phi_{1}P|\mathbf{h}_{1}\textbf{v}_{1}|^{2}}{\phi_{2}P|\mathbf{h}_{1}\mathbf{v}_{2}|^{2}+\sigma_{1}^{2}}= \frac{\phi_{1}P|\mathbf{h}_{2}\textbf{v}_{1}|^{2}}{\phi_{2}P|\mathbf{h}_{2}\mathbf{v}_{2}|^{2}+\sigma_{2}^{2}}
\end{align}
is always guaranteed in the solution to the optimization problem $\textbf{P1}$. In what follows, we prove \eqref{Rate-user-2-xishu-2} by contradiction. We first assume that
\begin{align}\label{Rate-user-2-xishu-proof1}
\frac{\phi_{1}P|\mathbf{h}_{1}\textbf{v}_{1}|^{2}}{\phi_{2}P|\mathbf{h}_{1}\mathbf{v}_{2}|^{2}+\sigma_{1}^{2}}> \frac{\phi_{1}P|\mathbf{h}_{2}\textbf{v}_{1}|^{2}}{\phi_{2}P|\mathbf{h}_{2}\mathbf{v}_{2}|^{2}+\sigma_{2}^{2}}
\end{align}
holds in the solution to the optimization problem $\textbf{P1}$. As per \eqref{Rate-user-2-xishu}, following  \eqref{Rate-user-2-xishu-proof1} we have
\begin{align}\label{C1-capacity_proof}
\gamma_{u1}= \frac{\phi_{1}P|\mathbf{h}_{2}\textbf{v}_{1}|^{2}}{\phi_{2}P|\mathbf{h}_{2}\mathbf{v}_{2}|^{2}+\sigma_{2}^{2}}.
\end{align}
Based on \eqref{beamforming-1}, in this case we can decrease $\beta_1$ in order to increase $\gamma_{u1}$ by slightly increasing the right-hand-side of \eqref{Rate-user-2-xishu-proof1} while decreasing its left-hand-side. This leads to the increase in the SSR (i.e., $R_s$), which is given in \eqref{SSR-1}, which is contradict to the assumption that \eqref{Rate-user-2-xishu-proof1} is guaranteed in the solution to the optimization problem $\textbf{P1}$. We have a similar argument for the suppose of
\begin{align}\label{Rate-user-2-xishu-proof2}
\frac{\phi_{1}P|\mathbf{h}_{1}\textbf{v}_{1}|^{2}}{\phi_{2}P|\mathbf{h}_{1}\mathbf{v}_{2}|^{2}+\sigma_{1}^{2}}< \frac{\phi_{1}P|\mathbf{h}_{2}\textbf{v}_{1}|^{2}}{\phi_{2}P|\mathbf{h}_{2}\mathbf{v}_{2}|^{2}+\sigma_{2}^{2}},
\end{align}
where we can increase the SSR $R_s$ by increasing $\beta_1$. As such, we complete the proof of Lemma~\ref{Lemma1}.
\end{IEEEproof}

Following Lemma~\ref{Lemma1}, for clarity in this work we write the SINR of $s_1$ as
\begin{align}\label{C1-capacity_result}
\gamma_{u1}=\frac{\phi_{1}P|\mathbf{h}_{1}\textbf{v}_{1}|^{2}}{\phi_{2}P|\mathbf{h}_{1}\mathbf{v}_{2}|^{2}+\sigma_{1}^{2}}.
\end{align}
Following \eqref{Rate-user-1}, \eqref{Rate-user-2}, and Lemma~\ref{Lemma1}, for given $\beta_1$ and $\beta_2$, the constraint $R_{um} \geq Q_m$ given in \eqref{YOHU-1a} can be rewritten as
\begin{align}\label{QoS-2}
&\phi_{1}\geq \frac{2^{Q_{1}}-1}{P|\mathbf{h}_{1}\textbf{v}_{1}|^{2}}\left(\phi_{2}P|\mathbf{h}_{1}\mathbf{v}_{2}|^{2}+\sigma_{1}^{2}\right),
\end{align}
and
\begin{align}\label{QoS-22}
&\phi_{2}\geq \frac{2^{Q_{2}}-1}{P|\mathbf{h}_{2}\textbf{v}_{2}|^{2}}\sigma_{2}^{2},
\end{align}
respectively.
\subsection{Secrecy Sum Rate with Sufficiently Large $N$}

Considering $N \rightarrow \infty$, we present an approximated but closed-form expression for the SSR (i.e., the objective function in the optimization problem $\textbf{P1}$) in the following theorem.

\begin{proposition}\label{Proposition 1}
As $N \rightarrow \infty$ with $N \gg K$, the SSR given in \eqref{SSR-1} can be approximated as
 \begin{align}\label{RS-MAX-approximate-01}
\widetilde{R}_{s}=& \left(1+\frac{((N-1)\beta_1+1)\phi_1\rho_{u1}}{\phi_{2}\rho_{u1}+1}\right)\left(1+((N-1)\beta_2+1)\phi_{2}\rho_{u2}\right)\notag\\
&+\log_{2}\left(\left(1+(1-\phi_{1}-\phi_{2})\rho_{e}\right)^{K}\right)-K\log_{2}(1+\rho_{e}),
\end{align}
where $\rho_{u1} = {P\delta_{1}^{2}}/{\sigma_{1}^{2}}$, $\rho_{u2} = {P\delta_{2}^{2}}/{\sigma_{2}^{2}}$, and $\rho_{e} = {P\delta_{e}^{2}}/{\sigma_{e}^{2}}$ are the average SNRs of the BS-User 1,  BS-User 2, and  BS-Eve links, respectively.
\end{proposition}

\begin{IEEEproof}
The proof is presented in Appendix~\ref{app1}.
\end{IEEEproof}

In the remaining of this work, we use the approximated SSR (i.e., $\widetilde{R}_{s}$) instead of $R_s$ as our objective function, since we focus on the scenario with a sufficiently large number of transmit antennas.

\begin{lemma}\label{Lemma 2}
As $N \rightarrow \infty$, the constraints in \eqref{QoS-2} and \eqref{QoS-22} can be rewritten as
\begin{align}\label{Constraints-P2-infty}
\phi_{1}\geq\frac{2^{Q_{1}}-1}{((N-1)\beta_1+1)\rho_{u1}}(1+\phi_{2}\rho_{u1}),
 \end{align}
and
\begin{align}\label{Constraints-P2-infty-1}
\phi_{2} \geq  \frac{2^{Q_{2}}-1}{((N-1)\beta_2+1)\rho_{u2}},
 \end{align}
 respectively.
\end{lemma}

\begin{IEEEproof}
The  proof of Lemma~\ref{Lemma 2} follows similar arguments as
that of Proposition~\ref{Proposition 1} and thus is omitted here.
\end{IEEEproof}

Based on Proposition~\ref{Proposition 1} and Lemma~\ref{Lemma 2}, We focus on the optimization problem $\mathbf{P2}$, instead of $\mathbf{P1}$, in the remaining of this work. Specifically, $\mathbf{P2}$ is expressed as
 \begin{align}\label{YOHU-9-modify}
&\hspace{-4mm}\mathbf{P2}:\operatorname*{max}\limits_{\phi_{1},\phi_{2},\phi_e,\beta_{1},\beta_{2}} \widetilde{R}_{s}\\ \label{YOHU-9-modify-2}
&{\rm  s.t.}\hspace{4mm}\phi_{1}\geq\frac{(2^{Q_{1}}-1)(1+\phi_{2}\rho_{u1})}{((N-1)\beta_1+1)\rho_{u1}}, \\
&\hspace{9mm} \phi_{2} \geq  \frac{2^{Q_{2}}-1}{((N-1)\beta_2+1)\rho_{u2}},\\
&\hspace{9mm} \phi_1 + \phi_2 + \phi_e = 1,  \\
&\hspace{9mm} 0 \leq \beta_m \leq 1, ~m \in \{1, 2\}.
     \end{align}

We will tackle  the optimization problem $\mathbf{P2}$ in the following section.

\section{Power Allocation and Beamforming Design \\in MISOME NOMA Systems}

In this section, we first solve $\mathbf{P2}$ for given values of $\beta_{1}$  and $\beta_{2}$, where we  recast the  SSR maximization as a two-level optimization framework that involves a one-dimensional numerical search. Then, we  analytically determine the optimal power allocation in the high-SNR regime for fixed $\beta_{1}$  and $\beta_{2}$. Finally, we provide the method of obtaining the optimal $\beta_{1}$  and $\beta_{2}$.

\subsection{Optimal Power Allocation for Given $\beta_1$ and $\beta_2$}

For given $\beta_1$ and $\beta_2$, the optimization problem $\mathbf{P2}$ can be rewritten as
    \begin{align}\label{YOHU-9-modify-0}
\mathbf{P3}:\operatorname*{max}\limits_{\phi_{1},\phi_{2},\phi_e} &\widetilde{R}_s(\beta_{1},\beta_{2})=\operatorname*{max}\limits_{\phi_{e}} \bigg\{\log_{2}\left(1+\phi_{e}\rho_{e}\right)^{K}\notag\\
&\hspace{0mm}+\operatorname*{max}\limits_{\phi_{1},\phi_{2},}\log_{2}\left[\left(1+\frac{c_1\phi_1}{1+\phi_2\rho_{u1}}\right)(1+c_2\phi_2)\right]\bigg\}\notag\\
&-K\log_{2}(1+\rho_{e})\\ \label{YOHU-9-modify-2}
{\rm  s.t.}&\hspace{3mm} \phi_{e}= 1-\phi_{1}-\phi_{2}, \\ \label{YOHU-9-modify-3}
&\hspace{3mm}\phi_{1}\geq\frac{1}{c_1}(2^{Q_{1}}-1)(1+\phi_{2}\rho_{u1}), \\
&\hspace{3mm} \phi_{2} \geq  \frac{1}{c_2}(2^{Q_{2}}-1),
     \end{align}
where  $\widetilde{R}_s(\beta_{1},\beta_{2})$ denotes $\widetilde{R}_{s}$ given in \eqref{YOHU-9-modify} for given $\beta_1$ and $\beta_2$, $c_1=(1+(N-1)\beta_1)\rho_{u1}$, and $c_2=(1+(N-1)\beta_2)\rho_{u2}$. Due to the high complexity of the objective function in the optimization problem $\mathbf{P3}$, we solve it in the following two steps. In the first step, for a given power allocation coefficient $\phi_e$, we obtain closed-form expressions for the optimal values of the power allocation coefficients $\phi_{1}$ and $\phi_{2}$, which are functions of $\phi_{e}$. In the second step, we adopt a one-dimensional numerical search to determine the optimal value of $\phi_e$, which leads to the optimal power allocation for given $\beta_1$ and $\beta_2$.

In the first step of solving the optimization problem $\mathbf{P3}$, we tackle the following optimization problem for a given $\phi_e$:
\begin{align}\label{YOHU-9-modify-inner}
\mathbf{P4}:& \operatorname*{max}\limits_{\phi_{1},\phi_{2}}F(\phi_{1},\phi_{2})\\ \label{YOHU-9-modify-inner-1}
{\rm  s.t.}&\hspace{3mm} \phi_{1}+\phi_{2}= 1-\phi_{e},\\ \label{YOHU-9-modify-inner-2}
&\hspace{3mm} \phi_{1}\geq\frac{1}{c_1}(2^{Q_{1}}-1)(1+\phi_{2}\rho_{u1}), \\ \label{YOHU-9-modify-inner-3}
& \hspace{3mm} \phi_{2} \geq   \frac{1}{c_2}(2^{Q_{2}}-1),
\end{align}
where
\begin{align}\label{YOHU-inner-define}
F(\phi_{1},\phi_{2})=\log_{2}\left[\left(1+\frac{c_1\phi_1}{1+\phi_2\rho_{u1}}\right)(1+c_2\phi_2)\right].
\end{align}
We note that the feasible range of $\phi_e$ is $0\leq\phi_{e}\leq\frac{P-P_{min}}{P}$, where we recall that $P_{min}$ is the minimum transmit power that guarantees the QoS constraints at the two legitimate users. We also note that due to the constraint given in \eqref{YOHU-9-modify-inner-1} the only parameter to optimize in $\mathbf{P4}$ is $\phi_1$ or $\phi_2$. Here, we take $\phi_2$ as the parameter to optimize. Then, we have the following lemma to facilitate solving the optimization problem $\mathbf{P4}$.

\begin{lemma}\label{Lemma 3}
For $\phi_{1}+\phi_{2}= 1-\phi_{e}$, the objective function in $\mathbf{P4}$, i.e., $F(\phi_{1},\phi_{2})$ given in \eqref{YOHU-inner-define}, is a concave function of $\phi_{2}$.
\end{lemma}
\begin{IEEEproof}
The proof is presented in Appendix~\ref{app2}.
\end{IEEEproof}

Following Lemma~\ref{Lemma 3}, the solution to the optimization problem $\mathbf{P4}$ is given in the following theorem.

\begin{theorem}\label{Theorem 1}
For a given feasible $\phi_e$, the optimal values of $\phi_1$ and $\phi_2$ for the optimization problem $\mathbf{P4}$ are derived as functions of $\phi_e$, given by
\begin{align}\label{opti-2}
&\phi_{2}^{\dag}(\phi_{e})\!=\!\left\{\begin{array}{ll}
\mu_{0}, & \text{when} ~ \mu_{1}\leq \mu_0\leq \mu_{2},\\
\mu_{1}, & \text{when} ~\mu_0 < \mu_1,\\
\mu_{2}, & \text{when} ~\mu_0 > \mu_2,\\
\end{array}
\right.\\
&\phi_{1}^{\dag}(\phi_{e})=1-\phi_{e}-\phi_{2}^{\dag}(\phi_{e}),
\end{align}
\end{theorem}
where
\begin{align}
&\mu_0=\frac{\sqrt{c_2^2c_3^2-c_2c_3\rho_{u1}c_4}-c_2c_3}{c_2c_3\rho_{u1}}, \\
&\mu _{1}=\frac{1}{c_2}(2^{Q_{2}}-1),\\
&\mu_{2}=\frac{1-\phi_{e}-\frac{1}{c_1}(2^{Q_{1}}-1)}{1+\frac{(2^{Q_1}-1)}{(N-1)\beta_1+1}}.
\end{align}

\begin{IEEEproof}
Based on Lemma~\ref{Lemma 3}, the optimal value of $\phi_{2}$ that maximizes the objective function in $\mathbf{P4}$ without considering the constraints given in \eqref{YOHU-9-modify-inner-2} and \eqref{YOHU-9-modify-inner-3} is the one that guarantees $\frac{\partial F(\phi_{1},\phi_{2})}{\partial\phi_{2}}=0$ (i.e., $G(\phi_{2})^{'}=0$ in \eqref{G-derivative-1} of Appendix~\ref{app2}), which is given by (following Lemma~\ref{Lemma 3} again)
\begin{align}
\mu_0 = \frac{\sqrt{c_2^2c_3^2-c_2c_3\rho_{u1}c_4}-c_2c_3}{c_2c_3\rho_{u1}}.
\end{align}
Substituting \eqref{YOHU-9-modify-inner-1} into \eqref{YOHU-9-modify-inner-2}, the constraints given in  \eqref{YOHU-9-modify-inner-2} and \eqref{YOHU-9-modify-inner-3} can be rewritten as the constraints on $\phi_2$, given by
\begin{align}\label{Constraint-phi2}
\mu_1 \leq \phi_{2}\leq \mu_2.
\end{align}
If $\mu_0$ satisfies the constraints given in \eqref{Constraint-phi2}, we can directly conclude $\phi_2^{\dag}(\phi_e) = \mu_0$. Otherwise, we have the following two cases. For $\mu_0 < \mu_1$, we have $\phi_2^{\dag}(\phi_e) = \mu_1$. This is due to the fact that, as we proved in Lemma~\ref{Lemma 3}, the objective function $F(\phi_{1},\phi_{2})$ is a concave function of $\phi_2$ and $\mu_0$ is the value of $\phi_2$ that maximizes $F(\phi_{1},\phi_{2})$, which leads to the fact that when $\mu_0 < \mu_1$ the objective function $F(\phi_{1},\phi_{2})$ monotonically decreases with $\phi_2$ for $\mu_1 \leq \phi_{2}\leq \mu_2$. Following a similar argument, we have $\phi_2^{\dag}(\phi_e) = \mu_2$ when $\mu_0 > \mu_2$. This completes the proof of Theorem~\ref{Theorem 1}.
\end{IEEEproof}

Following Theorem~\ref{Theorem 1}, in the second step of solving the optimization problem $\mathbf{P3}$ we have to solve a univarivate optimization problem with respect to  $\phi_{e}$, which is given by
\begin{align}\label{YOHU-9-modify-1}
\mathbf{P5}:\operatorname*{max}\limits_{0 \leq \phi_{e} \leq (P\!-\!P_{min})/P} \bigg[&\log_{2}\left(1\!+\!\phi_{e}\rho_{e}\right)^{K}\!+\!\operatorname*F(\phi_{1}^{\ast}(\phi_{e}),\phi_{2}^{\ast}(\phi_{e}))-K\log_{2}(1+\rho_{e})\bigg].
\end{align}
We note that the optimization problem $\mathbf{P5}$ is identical to $\mathbf{P3}$. For the optimization problem $\mathbf{P5}$, we can perform a one-dimensional numerical search over $0 \leq \phi_{e} \leq (P\!-\!P_{min})/P$ to determine the optimal value of $\phi_e$, i.e., $\phi_{e}^{\ast}$. Then, substituting $\phi_{e}^{\ast}$ into Theorem~\ref{Theorem 1} we can obtain the optimal values of $\phi_{1}$ and $\phi_{2}$, which are denoted by $\phi_{1}^{\ast}$ and $\phi_{2}^{\ast}$, respectively. So far, we have solved the optimization problem $\mathbf{P3}$ with the aid of a one-dimensional numerical search, which determines the optimal power allocation strategy for given $\beta_1$ and $\beta_2$. In order to reduce the complexity of determining the optimal power allocation and provide some insights based on analysis, in the following subsection we focus on analytically determining the optimal power allocation in the high-SNR regime.

\subsection{Optimal Power Allocation in High-SNR Regime}

In the high-SNR regime, i.e., as $\rho_{u1} \rightarrow \infty$ and $\rho_{u2} \rightarrow \infty$, following \eqref{RS-MAX-CB}, $R_{u1}+R_{u2}$ can be further approximated as
\begin{align}\label{Cb-Hihg-SNR}
R_{u1}+R_{u2}&=\log_{2}\left(1+\phi_{2}N\beta_2\rho_{u2}+\phi_{2}(1-\beta_2)\rho_{u2}\right)\notag\\
&\hspace{6mm}+\log_{2}\left(1+\frac{\phi_{1}\rho_{u1}+\phi_1(N-1)\beta_1\rho_{u1}}{\phi_{2}\rho_{u1}+1}\right)\notag\\
&=\log_{2}\left(c_2\phi_2\right)+\log_{2}\left(\frac{\phi_{1}((N-1)\beta_1+1)\rho_{u1}}{\phi_{2}\rho_{u1}}\right)\notag\\
&=\log_{2}\left(((N-1)\beta_{1}+1)c_2\phi_{1}\right).
\end{align}

Noting $\phi_1 + \phi_2 + \phi_e = 1$ and following \eqref{Cb-Hihg-SNR}, for given $\beta_{1}$  and $\beta_{2}$ the optimization problem $\mathbf{P2}$ can be rewritten as
\begin{align}\label{YOHU-9-modify-2}
\mathbf{P6}:\operatorname*{max}\limits_{\phi_{1},\phi_{2}} \widetilde{R}_{s}(\beta_{1},\beta_{2})&=\operatorname*{max}\limits_{\phi_{1},\phi_{2}} \bigg[\log_{2}\left(\phi_{1}\left(1+(1-\phi_{1}-\phi_{2})\rho_{e}\right)^{K}\right)\notag\\
&\hspace{10mm}+\log_{2}\left(((N-1)\beta_{1}+1)c_2\right)-K\log_{2}(1+\rho_{e})\bigg]\\
&{\rm  s.t.}\hspace{3mm}\phi_{1}\geq\frac{1}{c_1}(2^{Q_{1}}-1)(1+\phi_{2}\rho_{u1}), \label{YOHU-9-modify-3}\\
&\hspace{8mm} \phi_{2} \geq  \frac{1}{c_2}(2^{Q_{2}}-1).\label{YOHU-9-modify-4}
\end{align}
The solution to the optimization problem $\mathbf{P6}$ is presented in the following theorem.

\begin{theorem}\label{Theorem 2}
In the high SNR regime, i.e., as $\rho_{u1} \rightarrow \infty$ and $\rho_{u2} \rightarrow \infty$, the optimal power allocation coefficients, which are solutions to the optimization problem $\mathbf{P6}$, are derived as
 \begin{align}\label{Oprimal-coff-3}
&\phi_{1}^{\ast}=\left\{\begin{array}{ll}
\frac{1+\rho_{e}-\gamma_1\rho_{e}}{(K+1)\rho_{e}}, \hspace{-2.5mm} &\text{when} \: \gamma_{0} \leq \frac{1+\rho_{e}-\gamma_{1}\rho_{e}}{(K+1)\rho_{e}}\leq 1-\gamma_{1},\\
1-\gamma_{1}, &\text{when} \: \frac{1+\rho_{e}-\gamma_{1}\rho_{e}}{(K+1)\rho_{e}} > 1-\gamma_{1},\\
\gamma_0, & \text{when} \: \frac{1+\rho_{e}-\gamma_{1}\rho_{e}}{(K+1)\rho_{e}} < \gamma_{0},\\
\end{array}
\right.\\
&\phi_{2}^{\ast}=\gamma_1,\\
&\phi_{e}^{\ast}=1-\phi_{1}^{\ast}-\phi_{2}^{\ast},\end{align}
where $\gamma_{0}=\frac{1}{c_1}(2^{Q_{1}}-1)(1+\gamma_{1}\rho_{u1})$ and $\gamma_{1}= \mu_1$.
\end{theorem}

\begin{IEEEproof}
As per \eqref{YOHU-9-modify-2}, the objective function in $\mathbf{P6}$, i.e., the SSR $\widetilde{R}_{s}(\beta_{1},\beta_{2})$, monotonically decreases with $\phi_2$. Noting the constraints given in \eqref{YOHU-9-modify-3} and \eqref{YOHU-9-modify-4}, we  conclude that the optimal value of $\phi_2$ is the one that guarantees the equality in \eqref{YOHU-9-modify-4}, since decreasing $\phi_2$ makes the constraint given in \eqref{YOHU-9-modify-3} be guaranteed more easily.
After obtaining the optimal value of $\phi_2$ (i.e., $\phi_{2}^{\ast}=\gamma_{1}$), the optimization problem $\mathbf{P6}$ can be rewritten as
\begin{align}\label{YOHU-9-modify-4-1}
\mathbf{P7}:\operatorname*{max}\limits_{\phi_{1}} \widetilde{R}_{s}(\beta_{1},\beta_{2})&=\operatorname*{max}\limits_{\phi_{1}} \log_{2}\bigg[\left(\phi_{1}\left(1+(1-\phi_{1}-\gamma_1)\rho_{e}\right)^{K}\right)\notag\\
&\hspace{10mm}+\log_{2}(((N-1)\beta_{1}+1)c_2)-K\log_{2}(1+\rho_{e})\bigg]\\ \label{YOHU-9-modify-51}
{\rm  s.t.}&\hspace{3mm} \gamma_0 \leq\phi_{1}\leq 1-\gamma_1,
\end{align}
where the constraint given in \eqref{YOHU-9-modify-51} comes from \eqref {YOHU-9-modify-3}, \eqref{YOHU-9-modify-4}, and the consideration of $\phi_1 + \phi_2 + \phi_e = 1$ and $0\leq \phi_e$.

In the following, we first maximize the objective function in $\mathbf{P7}$ given in \eqref{YOHU-9-modify-4-1} without considering the constraint of \eqref{YOHU-9-modify-51}, which is presented in the following lemma.
\begin{lemma}\label{Lemma 4}
The term of $\phi_{1}\left(1+(1-\phi_{1}-\gamma_1)\rho_{e}\right)^{K}$ in the objective function of $\mathbf{P7}$ (i.e., the first term given in \eqref{YOHU-9-modify-4-1}) is maximized over $\phi_1$ when $\phi_{1}=\frac{1+\rho_{e}-\gamma_1\rho_{e}}{(K+1)\rho_{e}}$.
\end{lemma}
\begin{IEEEproof}
The proof is presented in Appendix~\ref{app3}.
\end{IEEEproof}
Based on \eqref{YOHU-9-modify-4-1}, we find that the second and third terms in the objective function of $\mathbf{P7}$ are not functions of $\phi_{1}$. Since $\log x$ is an increasing function of $x$, the value of $\log x$ is maximized when $x$ is maximized. As such, in order to maximize the objective function in $\mathbf{P7}$ (i.e.,  \eqref{YOHU-9-modify-4-1}) without considering the constraint of \eqref{YOHU-9-modify-51}, following Lemma~\ref{Lemma 4}, we have $\phi_{1}=\frac{1+\rho_{e}-\gamma_1\rho_{e}}{(K+1)\rho_{e}}$.

Now, we consider the constraint of \eqref{YOHU-9-modify-51} in $\mathbf{P7}$. Specifically, if $ \frac{1+\rho_{e}-\gamma_1\rho_{e}}{(K+1)\rho_{e}}$ satisfies the constraints given in \eqref{YOHU-9-modify-51}, we can directly conclude $\phi_1^{\ast}=\frac{1+\rho_{e}-\gamma_1\rho_{e}}{(K+1)\rho_{e}}$. Otherwise, we have the following two cases, which directly follow from the proof of Lemma~\ref{Lemma 4}. Specifically, if $\frac{1+\rho_{e}-\gamma_{1}\rho_{e}}{(K+1)\rho_{e}} < 1-\gamma_{1}$, we have $\phi_1^{\ast}=1-\gamma_1$. Otherwise,  we have $\phi_1^{\ast}=\gamma_0$. Finally, we can obtain the optimal power allocation coefficient for AN as $\phi_{e}^{\ast}=1-\phi_{1}^{\ast}-\phi_{2}^{\ast}$. This completes the proof of Theorem~\ref{Theorem 2}.
\end{IEEEproof}
Following Theorem~\ref{Theorem 2}, we note that $\phi_2^{\ast} P$ is a fixed value regardless of the total transmit power $P$ in the high-SNR regime, which is given by
\begin{align}\label{power2_optimal}
\phi_2^{\ast} P = \frac{{\sigma_{2}^{2}}}{(1+(N-1)\beta_2)\delta_{2}^{2}}(2^{Q_{2}}-1).
\end{align}
This indicates that for a downlink MISOME NOMA system, the optimal power allocation policy for maximizing the SSR is to use a fixed transmit power to User 2 in order to guarantee the equality in its QoS constraint and then allocate the remaining transmit power ($P-P_{\min}$) to User 1 or transmitting AN signals. This is different from the conclusion drawn in \cite{Y.Zhang2016}, which is that the extra transmit power is still allocated to User 2. As per \eqref{power2_optimal}, we also note that the transmit power allocated to User 2 (i.e., $\phi_2^{\ast} P$) decreases with $N$, which is not a function of $K$.

Following Theorem~\ref{Theorem 2}, for $\gamma_{0} \leq \frac{1+\rho_{e}-\gamma_{1}\rho_{e}}{(K+1)\rho_{e}}\leq 1-\gamma_{1}$, $\phi_e^{\ast}$ increases with  $K$ or $\rho_{e}$, while $\phi_{1}^{\ast}$ decreases with $K$ or $\rho_{e}$. This indicates that for a fixed total transmit power, we need to allocate more transmit power to AN when Eve's channel quality  becomes higher.

For $\frac{1+\rho_{e}-\gamma_{1}\rho_{e}}{(K+1)\rho_{e}}> 1-\gamma_{1}$, we have $\phi_e^{\ast} = 0$ as per Theorem~\ref{Theorem 2}, which indicates that under some specific conditions it is not necessary to transmit AN. We note that the value of $\frac{1+\rho_{e}-\gamma_{1}\rho_{e}}{(K+1)\rho_{e}}$ increases when $K$ or $\rho_e$ decreases. This indicates that as Eve's channel quality becomes lower the probability of the BS having to transmit AN decreases.

Following Theorem~\ref{Theorem 2}, for $\gamma_{0} > \frac{1+\rho_{e}-\gamma_{1}\rho_{e}}{(K+1)\rho_{e}}$ in the optimal power allocation we have that the transmit power allocated to User 1 and User 2 only guarantees the equality in their QoS constraints and then all the remaining transmit power is allocated to AN. The probability of this case increases with $K$ and $\rho_e$, for which Eve is a very strong eavesdropper.

\subsection{Optimization of Beamforming Parameters $\beta_{1}$ and $\beta_{2}$}

So far, we have presented the optimization of the power allocation coefficients for given beamforming parameters $\beta_{1}$ and $\beta_{2}$. In this subsection, we discuss the optimization framework of $\beta_{1}$ and $\beta_{2}$.

We note that Lemma~\ref{Lemma1} determines a one-to-one relationship between $\beta_{1}$ and $\beta_{2}$. Specifically, following \eqref{Rate-user-2-xishu} and considering $N \rightarrow\infty$, we have
\begin{align}\label{c1-min-capacity}
\gamma_{u1}^{1}(\beta_{1},\beta_{2})&\triangleq\frac{\phi_{1}P|\mathbf{h}_{1}\textbf{v}_{1}|^{2}}{\phi_{2}P|\mathbf{h}_{1}\mathbf{v}_{2}|^{2}+\sigma_{1}^{2}}\notag\\
&=\frac{\phi_{1}(N\beta_{1}+1)\rho_{u1}-\phi_1\beta_1\rho_{u1}}{\phi_{2}\rho_{u1}+1}\notag\\
&\approx\frac{\phi_{1}(N\beta_{1}+1)\rho_{u1}}{\phi_{2}\rho_{u1}+1},\\
\gamma_{u1}^{2}(\beta_{1},\beta_{2})&\triangleq\frac{\phi_{1}P|\mathbf{h}_{2}\textbf{v}_{1}|^{2}}{\phi_{2}P|\mathbf{h}_{2}\mathbf{v}_{2}|^{2}+\sigma_{2}^{2}}\notag\\
&=\frac{\phi_1N(1-\beta_1)\rho_{u2}+\phi_1\beta_1\rho_{u2}}{1+\phi_2(1+N\beta_2)\rho_{u2}}\notag\\
&\approx\frac{\phi_1N(1-\beta_1)\rho_{u2}}{1+\phi_2(1+N\beta_2)\rho_{u2}}.\label{c1-min-capacity-1}
\end{align}
Then, by setting $\gamma_{u1}^{1}(\beta_{1},\beta_{2}) = \gamma_{u1}^{2}(\beta_{1},\beta_{2})$ as per Lemma~\ref{Lemma1}, we can obtain the one-to-one relationship between $\beta_{1}$ and $\beta_{2}$. Thus,  the optimization problem $\mathbf{P2}$ can be rewritten as
\begin{align}\label{beta_1-beta-2}
\mathbf{P8}: &\max_{\beta_1,\beta_2} \;\;\widetilde{R}_{s}^{\ast}(\beta_{1},\beta_{2})\\
{\rm  s.t.}\; &\frac{\phi_{1}^{\ast}(N\beta_{1}+1)\rho_{u1}}{\phi_{2}^{\ast}\rho_{u1}+1}
=\frac{\phi_1^{\ast}N(1-\beta_1)\rho_{u2}}{1+\phi_2^{\ast}(1+N\beta_2)\rho_{u2}},\label{one-to-one}\\
&\hspace{0mm} 0 \leq \beta_m \leq 1, ~m \in \{1, 2\},
\end{align}
where $\widetilde{R}_{s}^{\ast}(\beta_{1},\beta_{2})$ is the maximum SSR achieved by the optimal power allocation for fixed $\beta_{1}$ and $\beta_{2}$.

As detailed in Section IV-A, the optimal power allocation for given $\beta_{1}$ and $\beta_{2}$ cannot be analytically determined with arbitrary SNRs, for which a one-dimensional numerical search is involved. As such, in the arbitrary SNR regime the optimization problem $\mathbf{P8}$ cannot be analytically solved. We note that a three-dimensional numerical search should be adopted to solve the optimization problem $\mathbf{P8}$, since the one-to-one relationship between $\beta_{1}$ and $\beta_{2}$ given in \eqref{one-to-one} depends on the optimal power allocation obtained from the one-dimensional numerical search. In order to reduce the complexity of solving optimization problem $\mathbf{P8}$ with a three-dimensional numerical search method, we next discuss how to efficiently solve it in the high-SNR regime.

In the high-SNR regime, following Lemma~\ref{Lemma1}, \eqref{c1-min-capacity}, and \eqref{c1-min-capacity-1}, we have
\begin{align}\label{xishu-2}
&\hspace{3mm}\frac{(N\beta_{1}+1)\phi_{1}\rho_{u1}}{\phi_{2}\rho_{u1}+1}=\frac{N\phi_1(1-\beta_1)\rho_{u2}}{1+(1+N\beta_2)\phi_2\rho_{u2}}\notag\\
&\Rightarrow \frac{(N\beta_{1}+1)\phi_{1}}{\phi_{2}}=\frac{N\phi_1(1-\beta_1)}{\phi_2(1+N\beta_2)}\notag\\
&\Rightarrow \beta_{2}=\frac{1}{1+N\beta_{1}}(1-\beta_{1}).
\end{align}
Then, the optimization problem $\mathbf{P8}$ can be rewritten as
\begin{align}\label{beta_1-beta-2}
\mathbf{P9}: &\max_{\beta_1,\beta_2} \;\;\widetilde{R}_{s}^{\ast}(\beta_{1},\beta_{2})\\
{\rm  s.t.}\; &\beta_{2}=\frac{1}{1+N\beta_{1}}(1-\beta_{1}),\label{one-to-one-high}\\
&\hspace{0mm} 0 \leq \beta_m \leq 1, ~m \in \{1, 2\},
\end{align}
which is identical to the optimization problem of $\mathbf{P2}$ in the high-SNR regime. We note that the optimization problem $\mathbf{P9}$ can be efficiently solved by a one-dimensional numerical search method, since as detailed in Theorem~\ref{Theorem 2} the optimal power allocation can be determined in closed-form expressions in the high-SNR regime, in which the one-to-one relationship given in \eqref{one-to-one-high} is also independent of the power allocation coefficients. As confirmed in our following numerical results, the achieved solution to $\mathbf{P9}$ is very close to the solution to $\mathbf{P8}$ and their resultant maximum SSRs are very similar to each other. This indicates that our proposed beamforming design can be optimized efficiently by a one-dimensional numerical search.

\section{Numerical Results}

In this section, we provide numerical results to examine the secrecy performances of the proposed NOMA-HB-AN scheme relative to two benchmark schemes. The first benchmark scheme is named as the NOMA-HB scheme, in which the beamforming design is the same as the proposed NOMA-HB-AN scheme, but no AN is transmitted by the BS. The second benchmark scheme is named as the NOMA-$\mathbf{h}_2$-AN scheme, which is proposed in \cite{L.Lv2018}. In the NOMA-$\mathbf{h}_2$-AN scheme, the beamforming vectors are set such as $\mathbf{v}_{1} = \mathbf{v}_{2}={\mathbf{h}_{2}}/{\|\mathbf{h}_{2}\|}$ and the AN signals are transmitted by the BS simultaneously. In this section, we set $\rho_{su}=\rho_{su2}=1.2\rho_{su1}$.

In Fig.~\ref{fig_sim2}, we plot the maximum SSRs achieved by the NOMA-HB-AN, NOMA-HB, and NOMA-$\mathbf{h}_2$-AN schemes with the optimal power allocation versus the number of antennas at base station (i.e., $N$). In this figure, the simulated SSR of the proposed NOMA-HB-AN scheme is obtained by performing Monte Carlo simulations over $10^{5}$ different channel realizations. We first observe that the analytical curve of the NOMA-HB-AN scheme accurately match the simulated one for different values of $N$, which confirms the high accuracy of the approximation adopted in our Proposition~1. In this figure, we also observe that the proposed NOMA-HB-AN scheme outperforms the
NOMA-$\mathbf{h}_2$-AN scheme. This demonstrates the effectiveness of the proposed beamforming design. We note that in this figure we set $\beta_{1}=0.05$ and $\beta_{2}=0.9$, which means that the performance gain of the proposed scheme over the NOMA-$\mathbf{h}_2$-AN scheme can be further improved by jointly optimizing $\beta_{1}$ and $\beta_{2}$. Furthermore, we observe that the proposed NOMA-HB-AN scheme outperforms the NOMA-HB scheme in terms of achieving a significantly higher maximum SSR, which shows the benefits of using AN-aided transmission schemes in enhancing physical layer security of NOMA systems. Finally, as expected we observe that the SSR increases with $N$.

 \begin{figure}[!t]
 \centering
 \includegraphics[width=4.3in, height=3.5in]{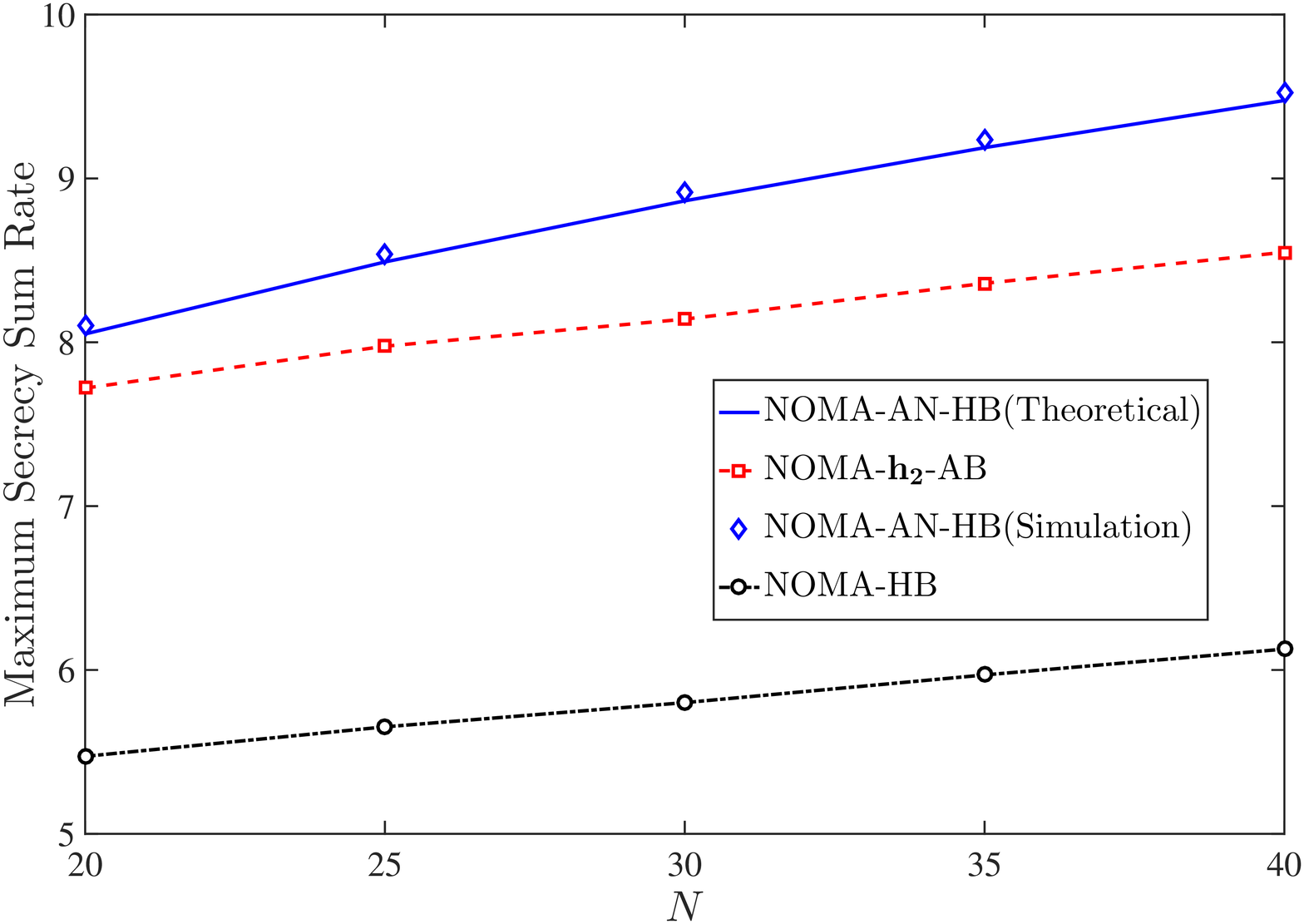}
 \caption{Maximum secrecy sum rate achieved by three different schemes versus the number of antennas at the BS $N$, where $\rho_{su}=20$ dB,  $\rho_{e}=10$ dB, $K=2$, $Q_{1}=5$ BPCU,  $Q_{2}=5.5$ BPCU, $\beta_{1}=0.05$, and $\beta_{2}=0.9$.}
 \label{fig_sim2}
 \end{figure}

\begin{figure}[!t]
 \centering
 \includegraphics[width=4.3in, height=3.5in]{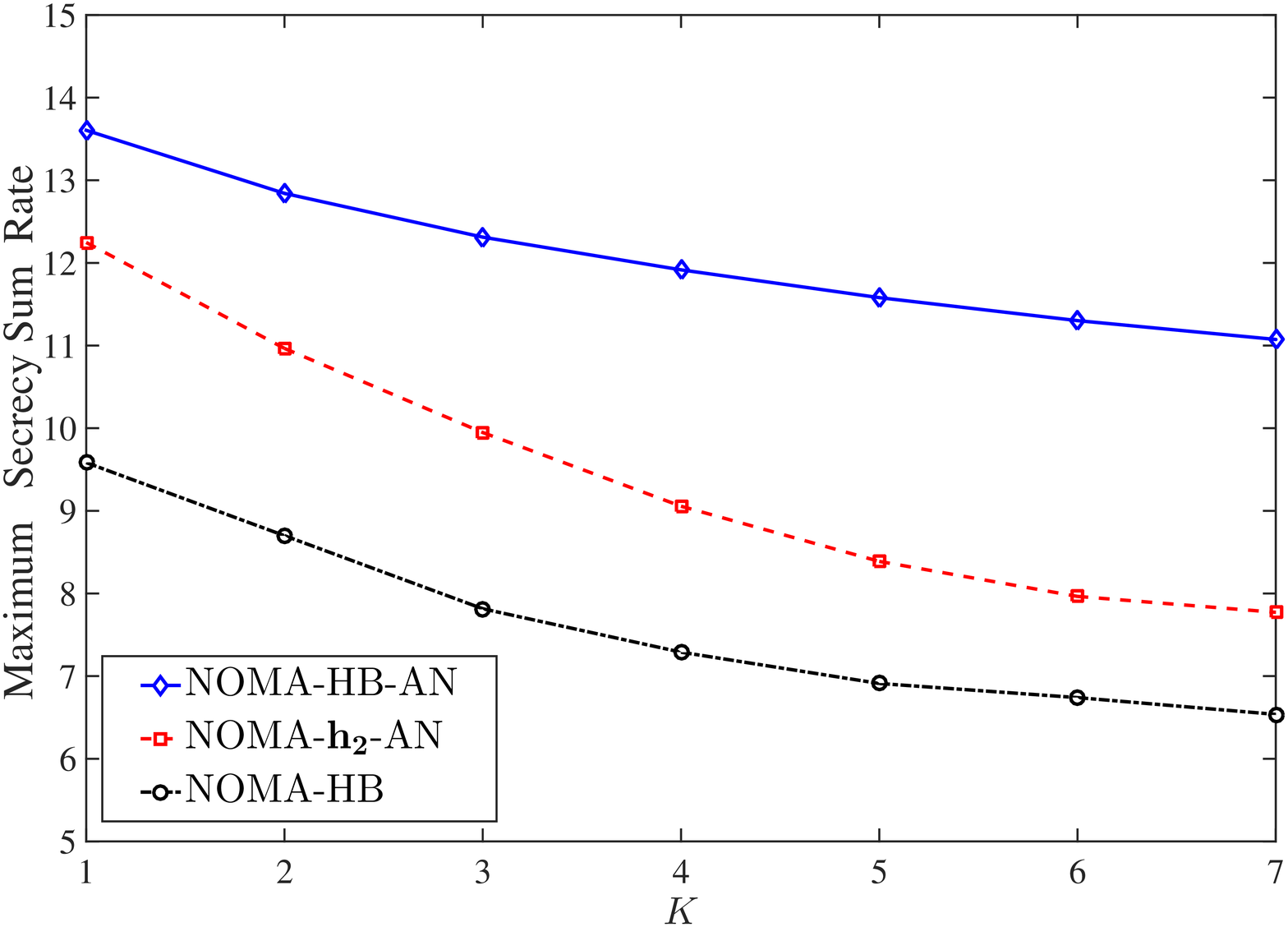}
 \caption{Maximum secrecy sum rate achieved by three different schemes versus the number of antennas at Eve $K$, where $Q_1=5$ BPCU,~$Q_2=5.5$ BPCU, $\rho_{su}=20$ dB,  $\rho_{e}=0$ dB, and $N=40$.}
 \label{fig_sim3}
 \end{figure}

In Fig.~\ref{fig_sim3}, we plot the maximum SSRs of the NOMA-HB-AN, NOMA-HB, and NOMA-$\mathbf{h}_2$-AN schemes versus the number of antennas at Eve (i.e., $K$). In this figure, the power allocation coefficients together with the values of $\beta_1$ and $\beta_2$ in the NOMA-HB-AN and NOMA-HB schemes have been optimized based on our conducted analysis in Section~III and Section~IV (e.g., Proposition~\ref{Proposition 1}, Theorem~\ref{Theorem 1}). The power allocation coefficients in the NOMA-$\mathbf{h}_2$-AN scheme have been optimized based on the analysis presented in \cite{L.Lv2018}. Again, in this figure we first observe that the proposed NOMA-HB-AN scheme significantly outperforms the two benchmark schemes, which again demonstrates the superiority of the proposed beamforming design. As expected, we observe that the achieved maximum SSRs decrease with $K$. We further observe that the performance gain of the proposed scheme over the NOMA-$\mathbf{h}_2$-AN scheme increases with $K$. This indicates that the advantage of the proposed scheme relative to the NOMA-$\mathbf{h}_2$-AN scheme becomes more dominant as $K$ increases, which shows the effectiveness of the proposed beamforming design increases with $K$.

 \begin{figure}[!t]
 \centering
  \includegraphics[width=4.3in, height=3.5in]{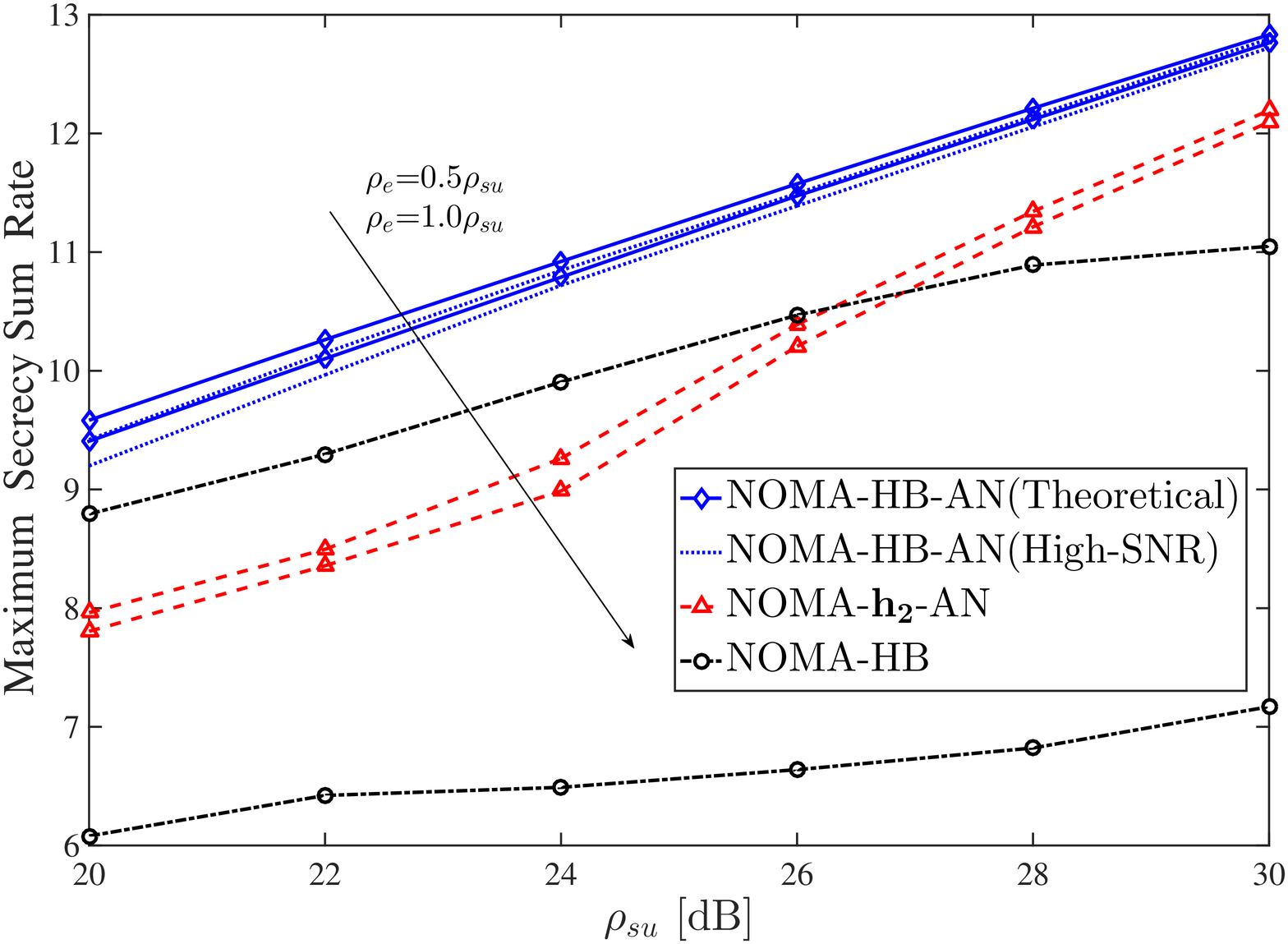}
 \caption{Maximum secrecy sum rate  achieved by three different schemes versus the average SNR $\rho_{su}$ with $\rho_{e}=0.5\rho_{su}$ and $\rho_{e}=\rho_{su}$, where  $N=30$, $Q_{1}=5$ BPCU, $Q_{2}=5.5$ BPCU, and $K=2$.}
 \label{fig_sim4}
 \end{figure}

In Fig.~\ref{fig_sim4}, we plot the maximum SSRs of the NOMA-HB-AN, NOMA-HB, and NOMA-$\mathbf{h}_2$-AN schemes versus the average SNR of the legitimate channels. We recall that we have set $\rho_{su}=\rho_{su2}=1.2\rho_{su1}$. In this figure, we first observe that, for the proposed NOMA-HB-AN scheme, the results achieved in the high-SNR regime (i.e., dashed curves) are very close to the exact results achieved for arbitrary SNRs (solid curves). We note that, in the high-SNR regime, the power allocation and the values of $\beta_1$ and $\beta_2$ are optimized by a one-dimensional numerical search with the aid of our Theorem~2, while, for arbitrary SNRs, the power allocation, $\beta_1$, and $\beta_2$ have to be optimized by a three-dimensional numerical search with the aid of our Theorem~1. As such, this observation demonstrates the usefulness of the asymptotic analysis in the high-SNR regime, i.e., the optimal power allocation together with the optimal $\beta_1$ and $\beta_2$ can be accurately approximated by those achieved in the high-SNR regime, which can significantly reduce the complexity of optimally designing the beamforming vectors and power allocation. As expected, we also observe that the proposed scheme outperforms the two benchmark schemes. Interestingly, we further observe that the NOMA-HB scheme can outperform the NOMA-$\mathbf{h}_2$-AN scheme for $\rho_{e}=0.5\rho_{su}$ and a relatively small $\rho_{su}$. This is due to the following two facts. First, the benefits of using AN decreases as $\rho_{e}$ decreases (i.e., Eve moves further from the BS). Second, the proposed hybrid beamforming design brings more benefits than using AN in the specific scenario. This observation again demonstrates the effectiveness of the our proposed novel beamforming design. Finally, in this figure we observe that when $\rho_{e}$ increases from $\rho_{e}=0.5\rho_{su}$ to $\rho_{e}=\rho_{su}$,  the secrecy performance of both the proposed NOMA-HB-AN scheme and the NOMA-$\mathbf{h}_2$-AN scheme only slightly decreases, but the NOMA-HB scheme suffers from a significant SSR degradation. This observation again shows the benefits of using the AN-aided transmission strategies to enhance physical layer security of NOMA systems.

 \begin{figure}[!t]
\centering
 \includegraphics[width=4.3in, height=3.5in]{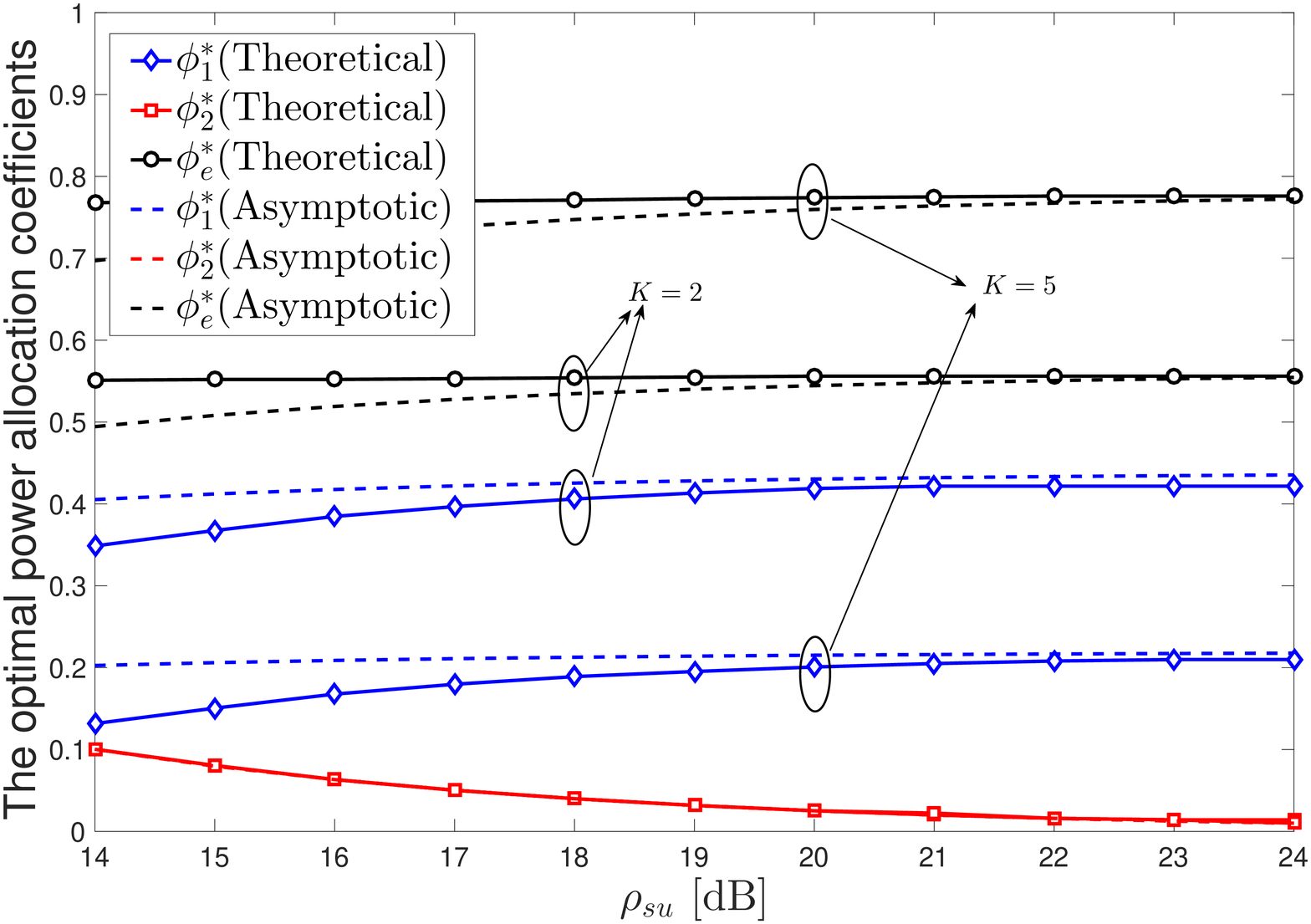}

\caption{The optimal power allocation coefficients versus the average SNR of the legitimate users' channels $\rho_{su}$  with  $K=2$ and  $K=5$, where $\rho_{e}=15$dB, $Q_{1}=2$ BPCU $Q_{2}=6$ BPCU, and  $N=20$.}
\label{fig_sim5}
\end{figure}

In Fig.~\ref{fig_sim5}, we  plot the optimal power allocation coefficients of  the NOMA-HB-AN scheme  versus the average SNR $\rho_{su}$ with  $K=2$ and  $K=5$. In this figure,  we first observe  that the power  allocation coefficients achieved  for arbitrary SNRs approach those achieved for high SNRs when $\rho_{su}$ increases, which again demonstrates the correctness and usefulness of our conducted asymptotic analysis in the high-SNR regime. Moreover, we observe that the the power allocation coefficients for User 1 and AN increase with $\rho_{su}$, while the power allocation coefficient for User 2 decreases when $\rho_{su}$ increases. This observation demonstrates that as the transmit power increases the remaining power (i.e., $P-P_{\min}$) is more likely allocated to either User 1 or used for transmitting AN signals in order to achieve the maximum SSR.
We further observe that  the power allocation coefficient $\phi_{e}^{\ast}$  increases with the number of antennas at Eve (i.e., $K$), illustrating the benefits of transmitting more AN signal to prevent eavesdropping attacks when the Eve's channel quality increases.  Finally, we observe that the power allocation coefficient $\phi_{2}^{\ast}$ is independent of the number of antennas at Eve.  These two observations can be explained by our  analysis in Section IV-B, i.e., Theorem~2 and its followed discussions.

\section{Conclusion}

In this work, we developed a novel beamforming design with the optimal power allocation to enhance physical layer security of a NOMA system. Specifically, we adopted two governing scalars to determine the beamforming vector, which can balance the SNR or SINR between the weak and strong users in the considered NOMA system. In order to demonstrate the benefits of our proposed beamforming design, we determined the optimal power allocation among information and AN signals by focusing on the asymptotic scenario with a large number of transmit antennas but an arbitrary SNR, based on which we also optimized the two governing scalars in the proposed beamforming design in order to maximize the SSR. In addition, we obtained an efficient solution to the optimization of the power allocation coefficients and the governing scalars in the high SNR regime, which is shown as a generic and near-optimal strategy. Our examination confirmed that our proposed NOMA-HB-AN scheme can significantly outperform the existing benchmark schemes, i.e., the NOMA-HB and NOMA-$\mathbf{h}_2$-AN schemes.

\appendices

\section{Proof of Proposition~\ref{Proposition 1}}\label{app1}
By substituting \eqref{Rate-user-1}, \eqref{Rate-user-2}, \eqref{Rate-Eva-1}, and \eqref{Rate-Eva-2} into \eqref{YOHU-1}, $R_{s}$ can be rewritten as
      \begin{align}\label{RS-MAX}
R_{s}=&\hspace{2mm} R_{u1}+R_{u2}-R_{e1}-R_{e2}\notag\\
     =&\hspace{2mm} \log_{2}\left(1+\frac{\phi_{2}P|\mathbf{h}_{2}\textbf{v}_{2}|^{2}}{\sigma_{2}^{2}}\right)+\log_{2}\left(1+\frac{\phi_{1}P|\mathbf{h}_{1}\textbf{v}_{1}|^{2}}{\phi_{2}P|\mathbf{h}_{1}\mathbf{v}_{2}|^{2}+\sigma_{1}^{2}}\right)\notag\\
                &-\log_{2}\det\left(\sigma_{e}^{2}\mathbf{I}_{K}+\frac{\phi_{2}P\mathbf{H}_{e}\mathbf{v}_{2}(\mathbf{H}_{e}\mathbf{v}_{2})^{H}}{\frac{\phi_eP}{N-2}\mathbf{H}_{e}\mathbf{V}_{N}(\mathbf{H}_{e}\mathbf{V}_{N})^{H}+\sigma_{e}^{2}\mathbf{I}_{K}}\right)\notag\\
                &-\log_{2}\det\left(\sigma_{e}^{2}\mathbf{I}_{K}+\frac{\phi_{1}P\mathbf{H}_{e}\mathbf{v}_{1}(\mathbf{H}_{e}\mathbf{v}_{1})^{H}}{P\phi_{2}\mathbf{H}_{e}\mathbf{v}_{2}(\mathbf{H}_{e}\mathbf{v}_{2})^{H}+\frac{\phi_eP}{N-2}\mathbf{H}_{e}\mathbf{V}_{N}(\mathbf{H}_{e}\mathbf{V}_{N})^{H}+\sigma_{e}^{2}\mathbf{I}_{K}}\right).
   \end{align}
To proceed, according to the law of large numbers, we have that, as $N \rightarrow \infty$,  the value of $|\mathbf{h}_{i}\mathbf{\hat{h}}_{i}|^{2}$ converges in probability to $N\delta_{i}^{2}$ ,  the value of both $|\mathbf{h}_{i}\mathbf{\hat{e}}|^{2}$ and $|\mathbf{h}_{i}\mathbf{\hat{h}}_{j}|^{2}$ with  $i\neq j$ converges in probability to $\delta_{i}^{2}$, and the value of both $\|\sqrt{\beta_{1}}\mathbf{\hat{h}}_{1} +\sqrt{ (1-\beta_{1})}\mathbf{\hat{h}}_{2}\|$ and $\|\sqrt{\beta_{2}}\mathbf{\hat{h}}_{2} + \sqrt{(1-\beta_{2})}\mathbf{\hat{e}}\|$ converges in probability to 1.  Then we have
\begin{align}\label{RS-MAX-CB}
R_{u1}+R_{u2}&= \log_{2}\left(1+\frac{\phi_{2}P|\mathbf{h}_{2}\textbf{v}_{2}|^{2}}{\sigma_{2}^{2}}\right)+\log_{2}\left(1+\frac{\phi_{1}P|\mathbf{h}_{1}\textbf{v}_{1}|^{2}}{\phi_{2}P|\mathbf{h}_{1}\mathbf{v}_{2}|^{2}+\sigma_{1}^{2}}\right)\notag\\
&\approx\log_{2}\left(1+\phi_{2}((N-1)\beta_2+1)\rho_{u2}\right)+\log_{2}\left(1+\frac{\phi_1((N-1)\beta_1+1)\rho_{u1}}{\phi_{2}\rho_{u1}+1}\right).
\end{align}
For the term $R_{e1}+R_{e2}$ in \eqref{YOHU-1}, we have
\begin{align}\label{RS-MAX-CE}
&R_{e1}+R_{e2}\notag\\
&=\log_{2}\det\left(\mathbf{I}_{K}+\frac{\phi_{2}\pi_{e}\mathbf{H}_{e}\mathbf{v}_{2}(\mathbf{H}_{e}\mathbf{v}_{2})^{H}}{\frac{\phi_{e}\pi_{e}}{N-2}\mathbf{H}_{e}\mathbf{V}_{N}(\mathbf{H}_{e}\mathbf{V}_{N})^{H}+\mathbf{I}_{K}}\right)\notag\\
                &\hspace{5mm}+\log_{2}\det\left(\mathbf{I}_{K}+\frac{\phi_{1}\pi_{e}\mathbf{H}_{e}\mathbf{v}_{1}(\mathbf{H}_{e}\mathbf{v}_{1})^{H}}{\phi_{2}\pi_{e}\mathbf{H}_{e}\mathbf{v}_{2}(\mathbf{H}_{e}\mathbf{v}_{2})^{H}+\frac{\phi_{e}\pi_{e}}{N-2}\mathbf{H}_{e}\mathbf{V}_{N}(\mathbf{H}_{e}\mathbf{V}_{N})^{H}+\mathbf{I}_{K}}\right)\notag\\
                &=\log_{2}\det\left(\mathbf{I}_{K}+\phi_{1}\pi_{e}\mathbf{H}_{e}\mathbf{v}_{1}(\mathbf{H}_{e}\mathbf{v}_{1})^{H}+\phi_{2}\pi_{e}\mathbf{H}_{e}\mathbf{v}_{2}(\mathbf{H}_{e}\mathbf{v}_{2})^{H}+\frac{\phi_{e}\pi_{e}}{N-2}\mathbf{H}_{e}\mathbf{V}_{N}(\mathbf{H}_{e}\mathbf{V}_{N})^{H}\right)\notag\\
                &\hspace{5mm}-\log_{2}\det\left(\mathbf{I}_{K}+\frac{\phi_{e}\pi_{e}}{N-2}\mathbf{H}_{e}\mathbf{V}_{N}(\mathbf{H}_{e}\mathbf{V}_{N})^{H}\right).
   \end{align}

In order to simplify \eqref{RS-MAX}, we present the following lemma (i.e., {Lemma}~\ref{lemma 3}) to facilitate our proof.
 \begin{lemma}\label{lemma 3}
As $N \rightarrow \infty$  with $N \gg K$, we have
 \begin{align}\label{RS-MAX-CE-2-1}
&\hspace{0mm}\det\bigg(\mathbf{I}_{K}+\phi_{1}\pi_{e}\mathbf{H}_{e}\mathbf{v}_{1}(\mathbf{H}_{e}\mathbf{v}_{1})^{H}+\phi_{2}\pi_{e}\mathbf{H}_{e}\mathbf{v}_{2}(\mathbf{H}_{e}\mathbf{v}_{2})^{H}+\frac{\phi_{e}\pi_{e}}{N-2}\mathbf{H}_{e}\mathbf{V}_{N}(\mathbf{H}_{e}\mathbf{V}_{N})^{H}\bigg)\notag\\
 & \!\leq\!\prod\limits_{k\!=\!1}^{K}  \bigg(1\!+\!\phi_{1}\pi_{e}[\mathbf{H}_{e}\mathbf{v}_{1}(\mathbf{H}_{e}\mathbf{v}_{1})^{H}]_{kk}\!+\!\phi_{2}\pi_{e}[\mathbf{H}_{e}\mathbf{v}_{2}(\mathbf{H}_{e}\mathbf{v}_{2})^{H}]_{kk}\!+\!\frac{(1\!-\!\phi_{1}\!-\!\phi_{2})\pi_{e}}{N\!-\!2}[\mathbf{H}_{e}\mathbf{V}_{N}(\mathbf{H}_{e}\mathbf{V}_{N})^{H}]_{kk}\bigg) \notag\\
 &\approx \left(1+\phi_{1}\rho_{e}+\phi_{2}\rho_{e}+(1-\phi_{1}-\phi_{2})\rho_{e}\right)^{K}\notag\\
 &=\left(1+\rho_{e}\right)^{K},
   \end{align}
and
\begin{align}\label{RS-MAX-CE-2-2}
\det\left(\mathbf{I}_{K}+\frac{\phi_{e}\pi_{e}}{N-2}\mathbf{H}_{e}\mathbf{V}_{N}(\mathbf{H}_{e}\mathbf{V}_{N})^{H}\right)
 \approx \prod\limits_{k=1}^{K} \left(1+\frac{\phi_e\pi_{e}}{N-2}[\mathbf{H}_{e}\mathbf{V}_{N}(\mathbf{H}_{e}\mathbf{V}_{N})^{H}]_{kk}\right)
=  \left(1+\phi_e\rho_{e}\right)^{K}.
\end{align}
\end{lemma}
The proof of {Lemma}~\ref{lemma 3} is given in \cite{S-HTsai2014} and thus it is omitted here. Based on \eqref{RS-MAX-CB}, \eqref{RS-MAX-CE}, and Lemma~\ref{lemma 3}, we can obtain \eqref{RS-MAX-approximate-01}, which completes the proof of Proposition~\ref{Proposition 1}.

\section{Proof of Lemma 3}\label{app2}
Substituting  \eqref{YOHU-9-modify-inner-2} into \eqref{YOHU-9-modify-inner}, we have
   \begin{align}\label{YOHU-inner-define-appl}
F(\phi_{1},\phi_{2})&=\log_{2}\left(1+\frac{c_1(1-\phi_2-\phi_e)}{1+\phi_2\rho_{u1}}\right)(1+c_2\phi_2)\triangleq G(\phi_{2}) .
     \end{align}
From \eqref{YOHU-inner-define-appl}, we can obtain  the first  derivative of  $G(\phi_{2})$ with respect to $\phi_2$ as  $G(\phi_{2})^{'} $,  which is give by
   \begin{align}\label{G-derivative-1}
 G(\phi_{2})^{'}&=\frac{\partial G(\phi_{2})}{\partial\phi_{2}}=\frac{1}{\ln(2)}\frac{c_2c_3\rho_{u1}\phi_2^2+2c_2c_3\phi_2+c_4}{(1+\phi_{2}\rho_{u1})^{2}},
     \end{align}
  where
  \begin{align}\label{c3}
  c_3&=\rho_{u1}-c_1,\\
  c_4&=(1+(1-\phi_e)c_1)(c_2-\rho_{u1})+\rho_{u1}-c_1.
  \end{align}
Furthermore, we can  obtain  the second derivative of  $G(\phi_{2})$ with respect to $\phi_2$ as  $G(\phi_{2})^{''}$,  which is give by
        \begin{align}\label{G-derivative-2}
& G(\phi_{2})^{''}=\frac{\partial G(\phi_{2})^{'}}{\partial\phi_{2}}=-\frac{1}{\ln(2)}\frac{c_1(2-\phi_e)((N-1)\beta_2\rho_{u2}+\rho_{u2}-\rho_{u1})}{(1+\phi_{2}\rho_{u1})^{3}}<0,
     \end{align}
due to the facts $\phi_e<1$ and $\rho_{u1}\leq\rho_{u2}$.
Following \eqref{G-derivative-2},  we find that $G(\phi_{2})$ is a concave function of $\phi_{2}$, which completes the proof of Lemma~\ref{Lemma 3}.

\section{Proof of Lemma 4}\label{app3}

Following \eqref{YOHU-9-modify-4-1}, we derive the first  derivative of  $\phi_{1}\left(1+(1-\phi_{1}-\gamma_1)\rho_{e}\right)^{K} \triangleq Q(\phi_{1})$ with respective to $\phi_{1}$ as
\begin{align}\label{Q-derivative-1}
Q(\phi_{1})^{'}&\triangleq\frac{\partial (\phi_{1}\left(1+(1-\phi_{1}-\gamma_1)\rho_{e}\right)^{K})}{\partial\phi_{1}}\notag\\
&=(\left(1+(1-\phi_{1}-\gamma_1)\rho_{e}\right)^{K-1}\left((\left(1+(1-\phi_{1}-\gamma_1)\rho_{e}\right)-K\phi_1\rho_e\right).
\end{align}
Following \eqref{Q-derivative-1} and noting $1-\phi_{1}-\gamma_1 \geq 0$, we have
\begin{align}\label{Q-derivative-1-0}
\phi_{1}=\phi_1^{\dag} \triangleq
\frac{1+\rho_{e}-\gamma_1\rho_{e}}{(K+1)\rho_{e}},
\end{align}
in order to guarantee $Q(\phi_{1})^{'}=0$.
As per \eqref{Q-derivative-1}, we can find that $Q(\phi_{1})^{'}>0$ for $\phi_1 < \phi_1^{\dag}$, which indicates that
the function of $Q(\phi_{1})$ is a monotonically increasing function of $\phi_1$ when $\phi_{1}>\phi_1^{\dag}$. We also find that $Q(\phi_{1})^{'}<0$ for $\phi_1 > \phi_1^{\dag}$, which shows that the function of $Q(\phi_{1})$ is a monotonically decreasing function of $\phi_1$ when $\phi_1 > \phi_1^{\dag}$. As such, we can conclude that $Q(\phi_{1})$ is maximized when $Q(\phi_{1})^{'}=0$, i.e., when \eqref{Q-derivative-1-0} is guaranteed.
This completes the proof of Lemma~\ref{Lemma 4}.


\begin{thebibliography}{10}

\bibitem{Y.Saito2013}
Y. Saito, A. Benjebbour, Y. Kishiyama, and T. Nakamura, ``System-level performance evaluation of downlink non-orthogonal multiple access (NOMA),'' in  \emph{Proc. IEEE PIMRC}, London, UK, Sep. 2013, pp. 611--615.

\bibitem{L.Dai2015}
L. Dai, B. Wang, Y. Yuan, S. Han, C.-L. I, and Z. Wang, ``Nonorthogonal
multiple access for 5G: Solutions, challenges, opportunities, and future research trends,'' \emph{IEEE Commun. Mag.}, vol. 53, no. 9, pp. 74--81, Sep. 2015.

\bibitem{Z.Ding2017JSAC}
 Z. Ding, X. Lei, G. K. Karagiannidis, R. Schober, J. Yuan, and V. K. Bhargava, ``A survey on non-orthogonal multiple access for 5G networks: Research challenges and future trends,''  \emph{IEEE J. Select. Areas Commun.}, vol. 35, no. 10, pp. 2181--2195, Oct. 2017.


 \bibitem{Y. Liu2017-1}
 Y. Liu, Z. Qin, M. Elkashlan, Z. Ding, A. Nallanathan, and L. Hanzo, ``Nonorthogonal multiple access for 5G and beyond,''  \emph{Proc. IEEE}, vol. 105, no. 12, pp. 2347--2381, Dec. 2017.

\bibitem{A.Benjebbovu2013}
 A. Benjebbovu, A. Li, Y. Saito, Y. Kishiyama, A. Harada, and T. Nakamura, ``System-level performance of downlink NOMA for future LTE enhancements,'' in  \emph{Proc. IEEE Globecom}, Dec. 2013, pp. 66--70.

\bibitem{J. Choi2016}
 J. Choi, ``On the power allocation for MIMO-NOMA systems with
layered transmissions,'' \emph{IEEE Trans. Wireless Commun.}, vol. 15, no. 5,
pp. 3226--3237, May 2016.

\bibitem{Y. Sun2017}
Y. Sun, D. W. K. Ng, Z. Ding, and R. Schober, ``Optimal joint power and
subcarrier allocation for full-duplex multicarrier non-orthogonal multiple
access systems,'' \emph{IEEE Trans. Commun.}, vol. 65, no. 3, pp. 1077--1091,
Mar. 2017.


\bibitem{X.Sun201802}
X. Sun, S. Yan, N. Yang, Z. Ding, C. Shen, and Z. Zhong, ``Short-packet downlink transmission with non-orthogonal multiple access,''  \emph{IEEE Trans. Wireless Commun.},  Early Access, Apr. 2018.
\bibitem{X.Sun201801}
X. Sun, N. Yang, S. Yan, Z. Ding,
D. W. K. Ng, C. Shen, and Z. Zhong, ``Joint beamforming and power allocation in
downlink NOMA multiuser MIMO networks,''  \emph{IEEE Trans. Wireless Commun.},  Early Access, Jun. 2018.

\bibitem{S.Yan2015}
S. Yan, N. Yang, G. Geraci, R. Malaney, and J. Yuan, ``Optimization
of code rates in SISOME wiretap channels,''  \emph{IEEE Trans. Wireless Commun.}, vol. 14, no. 11, pp. 6377--6388, Nov. 2015.

\bibitem{zou2016}
Y. Zou, J. Zhu, X. Wang, and L. Hanzo, ``A survey on wireless security: Technical challenges, recent advances, and future trends,'' \emph{Proc. IEEE}, vol. 104, no. 9, pp. 1727--1765, Sep. 2016.

\bibitem{T.-X.Zheng2014}
T.-X. Zheng, H.-M. Wang, J. Yuan, D. Towsley, and M. H. Lee, ``Multiantenna
transmission with artificial noise against randomly distributed
eavesdroppers,''  \emph{IEEE Trans. Commun.},  vol. 63, no. 11, pp. 4347--4362,
Nov. 2015.

\bibitem{Y.Hong2016}
Y. W. P. Hong, P. C. Lan, and C. C. J. Kuo, ``Enhancing physical-layer
secrecy in multi-antenna wireless systems: An overview of signal processing
approaches," \emph{ IEEE Signal Process. Mag.,} vol. 30, no. 5, pp. 29--40,
Sep. 2013.


\bibitem{YouhongFeng2017}
Y. Feng, Z. Yang, W.-P. Zhu, Q. Li, and B. Lv,
``Robust cooperative secure beamforming for simultaneous wireless information and power transfer in amplify-and-forward relay networks,''  \emph{IEEE Trans. Veh. Technol.},  vol. 66, no. 3, pp.
2354--2366, Mar. 2017.

\bibitem{Y. Feng201702}
 Y. Feng, S. Yan, Z. Yang, S. Yan, N. Yang, and W.-P. Zhu, `TAS-Based incremental hybrid decode-amplify-forward relaying for physical layer security enhancement,''  \emph{IEEE Trans. Commun.}, vol. 65, no. 9 pp. 3876--3891, Sep. 2017.


\bibitem{yang2015safeguarding}
N. Yang, L. Wang, G. Geraci, M. Elkashlan, J. Yuan, and M. Di Renzo, ``Safeguarding 5G wireless communication networks using physical layer security,'' \emph{IEEE Commun. Mag.}, vol. 53, no. 4, pp. 20-27, Apr. 2015.


\bibitem{N.Yang2016}
N. Yang, S. Yan, J. Yuan, R. Malaney, R. Subramanian, and I. Land, ``Artificial noise: Transmission optimization in multi-input single-output wiretap channels,'' \emph{IEEE Trans. Commun.},  vol. 63, no. 5, pp. 1771--1783, May 2015.



\bibitem{Z.Ding2017TC}
Z. Ding, Z. Zhao, M. Peng, and H. V. Poor, ``On the spectral efficiency and security enhancements of NOMA assisted
multicast-unicast streaming,'' \emph{IEEE Trans. Commun.}, vol. 65, no. 7, pp. 13151--3163, Jul. 2017.

\bibitem{Z.Qin2016}
Z. Qin, Y. Liu, Z. Ding, Y. Gao, and M. Elkashlan, ``Physical layer
security for 5G non-orthogonal multiple access in large-scale networks,''
in \emph{IEEE ICC}, May 2016, pp. 1--6.

\bibitem{Y.Zhang2016}
Y. Zhang, H.-M. Wang, Q. Yang, and Z. Ding, ``Secrecy sum rate
maximization in nonorthogonal multiple access,''  \emph{IEEE Commun. Lett.},
vol. 20, no. 5, pp. 930--933, May 2016.

\bibitem{F. Zhou2018}
F. Zhou,  Z. Chu,  H. Sun, R. Q. Hu,  and L. Hanzo, ``Artificial noise aided secure cognitive
beamforming for cooperative MISO-NOMA using
SWIPT,''  \emph{IEEE J. Sel. Areas Commun.},  Early Access,
Apr. 2018.

\bibitem{M. Jiang2017sp}
M. Jiang, Y. Li, Q. Zhang , Q. Li, and J. Qin, ``Secure beamforming in downlink MIMO
non-orthogonal multiple access networks,''  \emph{IEEE
Signal Process. Lett.}, vol. 24, no. 12, pp. 1852--1855, Aug. 2017.

\bibitem{M.Tian2017sp}
M. Tian, Q. Zhang, S. Zhao, Q. Li, and J. Qin, ``Secrecy sum rate optimization for downlink MIMO nonorthogonal multiple access systems,''  \emph{IEEE
Signal Process. Lett.}, vol. 24, no. 8, pp. 1113--1117, Aug. 2017.

\bibitem{B.He2017}
B. He, A. Liu, N. Yang, and V. K. N. Lau, ``On the design of secure non-orthogonal
multiple access systems,''  \emph{IEEE J. Sel. Areas Commun.}, vol. 35,
no. 10, pp. 2196--2206, Oct. 2018.

\bibitem{Y.Liu2017}
Y. Liu,  Z. Qin, M. Elkashlan,
Y. Gao, and L. Hanzo, ``Enhancing the physical layer security of
non-orthogonal multiple access in
large-scale networks,'' \emph{IEEE Trans. Wireless Commun.}, vol. 16, no. 3, pp. 1656--
1671, Dec. 2017.

\bibitem{L.Lv2018}
L. Lv,  Z. Ding, Q. Ni, and J. Chen, ``Secure MISO-NOMA transmission with
artificial noise,''  \emph{IEEE Trans. Veh. Technol.}, Early Access, Mar. 2018.


\bibitem{Y.Zhang2017}
Y. Zhang, H.-M. Wang, T.-X. Zheng, and Q. Yang, ``Energy-efficient transmission design in
non-orthogonal multiple access,''  \emph{IEEE  Trans. Veh. Technol.}, vol. 66, no. 3,  pp. 2852--2857, Mar. 2017.

\bibitem{Z.Chen2016}
 Z. Chen, Z. Ding, P. Xu, and X. Dai, ``Optimal precoding for a QoS optimization
problem in two-user MISO-NOMA downlink,''  \emph{IEEE Commun.
Lett.}, vol. 20, no. 6, pp. 1263--1266, Jun. 2016.

\bibitem{S.Yan2016}
S. Yan, X. Zhou, N. Yang, B. He, and T. D. Abhayapala,
``Artificial-noise-aided secure transmission in wiretap channels with transmitter-side
correlation,''  \emph{IEEE Trans. Wireless Commun.}, vol. 15, no. 12, pp. 8286--8297, Dec. 2016.

\bibitem{L.Fan2016}
S. Yan, N. Yang, R. Malaney, and J. Yuan, ``Transmit antenna selection with alamouti coding and power allocation in MIMO wiretap channels,''  \emph{IEEE Trans. Wireless Commun.}, vol. 13, no. 3, pp. 1656--1667, Mar. 2014.


\bibitem{Z.Ding2014}
Z. Ding, Z. Yang, P. Fan, and H. Poor, ``On the performance of nonorthogonalmultiple
access in 5G systems with randomly deployed users,''  \emph{
IEEE Signal Process. Lett.}, vol. 21, no. 12, pp. 1501--1505, Dec. 2014.


\bibitem{S. Goel2008}
S. Goel and R. Negi, ``Guaranteeing secrecy using artificial noise,''  \emph{
IEEE Trans. Wireless Commun.},  vol. 7, no. 6, pp. 2180--2189, Jun.
2008.
\bibitem{N.Li2016}
N. Li, X. Tao, H. Wu, Q. Cui, and J. Xu, ``Large-system analysis
of artificial-noise-assisted communication in the multiuser downlink:
Ergodic secrecy sum rate and optimal power allocation,''  \emph{IEEE Trans.
Veh. Technol.}, vol. 65, no. 9, pp. 7036--7050, Sep. 2016.

\bibitem{S-HTsai2014}
S.-H.  Tsai  and  H.  V.  Poor, ``Power  allocation  for  artificial-noise  secure
MIMO precoding systems,''  \emph{IEEE Trans. Signal Process.}, vol. 62, no. 13,
pp. 3479--3493, Jul. 2014.

\bibitem{N.Li2015}
 N. Li, X. Tao, and J. Xu, ``Artificial noise assisted communication in the
multiuser downlink: Optimal power allocation,''  \emph{IEEE Commun. Lett.},
vol. 19, no. 2, pp. 295--298, Feb. 2015.
\bibitem{Y. Sun2018}
Y.  Sun,  D. W. K. Ng, J. Zhu, and R. Schober, ``Robust and secure resource allocation for
full-duplex MISO multi-carrier NOMA systems,''  \emph{
IEEE Trans. Wireless Commun.},  Early Access, Jun. 2018.



\end{thebibliography}
\end{document}